\documentclass[a4paper,journal,twoside]{IEEEtran}
\ifCLASSINFOpdf
\else
\fi

\usepackage[cmex10]{amsmath}
\usepackage{amssymb}
\usepackage{mathtools}
\usepackage{epsfig}
\usepackage{epstopdf}
\usepackage{subfigure}
\usepackage{booktabs}
\usepackage{multirow}
\usepackage{color}
\usepackage{xcolor}
\usepackage{standalone}

\usepackage[noadjust]{cite}
\newsavebox{\ieeealgbox}

\usepackage{array}
\usepackage{mdwmath}
\usepackage{mdwtab}

\usepackage{stfloats}
\fnbelowfloat
\DeclareMathOperator{\RSNR}{SNR_{rec}}

\newcommand{\ACal}{\mathcal{A}}
\newcommand{\Ab}{\mathbf{A}}
\newcommand{\bb}{\mathbf{b}}
\newcommand{\Xb}{\mathbf{X}}
\newcommand{\xb}{\mathbf{x}}
\newcommand{\yb}{\mathbf{y}}
\newcommand{\Zb}{\mathbf{Z}}
\newcommand{\Xbh}{\widehat{\Xb}}
\newcommand{\XbF}{\widetilde{\Xb}}
\newcommand{\Xbd}{\Xb_{\delta}}
\newcommand{\Rbb}{\mathbb{R}}
\newcommand{\LOp}{\ACal: \Rbb^{n_1 \times n_2} \rightarrow \Rbb^m}
\newcommand{\nDef}{n=\min(n_1,n_2)}
\newcommand{\ICal}{\mathcal{I}}
\newcommand{\UDelmOne}{\lceil \Delta-1 \rceil}

\newcommand{\SigbXb}{\boldsymbol{\sigma}(\Xb)}
\newcommand{\SigbXbF}{\boldsymbol{\sigma}(\XbF)}
\newcommand{\SigbXbd}{\boldsymbol{\sigma}(\Xbd)}
\newcommand{\fdel}{f_{\delta}}
\newcommand{\Fdel}{F_{\delta}}
\newcommand{\sumn}{\sum_{i=1}^{n}}

\DeclareMathOperator{\rank}{rank}
\DeclareMathOperator{\trace}{trace}
\DeclareMathOperator{\vect}{vec}
\DeclareMathOperator{\mat}{mat}
\DeclareMathOperator*{\argmax}{argmax}
\DeclareMathOperator*{\argmin}{argmin}

\DeclareMathOperator{\nullS}{null}

\def\ROne_Col{black}
\def\RTwo_Col{black}
\def\RTre_Col{black}
\def\RFor_Col{black}
\def\MinEdt{black}
\def\FinEdt{black}

\begin{document}
\title{Recovery of Low-Rank Matrices under Affine Constraints via a Smoothed Rank Function}

\author{Mohammadreza~Malek-Mohammadi,~\IEEEmembership{Student,~IEEE}, Massoud~Babaie-Zadeh,~\IEEEmembership{Senior Member,~IEEE}, Arash Amini, and~Christian~Jutten,~\IEEEmembership{Fellow,~IEEE}% <-this % stops a space
\thanks{This work has been supported in part by the Iran Telecommunication Research Center (ITRC) under contract number 500/11307 {\color{\MinEdt}and Iran National Science Foundation under contract number 91004600.}}% <-this % stops a space
\thanks{M. Malek-Mohammadi, M. Babaie-Zadeh, and {\color{\MinEdt}A. Amini} are with the Electrical Engineering Department, Sharif University of Technology, Tehran 1458889694, Iran (e-mail:
m.rezamm@ieee.org; mbzadeh@yahoo.com; {\color{\MinEdt}aamini@sharif.edu}).}
\thanks{C. Jutten is with the GIPSA-Lab, Department of Images and Signals, University of Grenoble and Institut Universitaire de France, France (e-mail: Christian.Jutten@inpg.fr).}}

\maketitle

\begin{abstract}
In this paper, the problem of {\color{\MinEdt}matrix rank minimization} under affine constraints is addressed. The state-of-the-art algorithms can recover matrices with a rank much less than what is sufficient for the uniqueness of the solution of this optimization problem. We propose an algorithm based on a smooth approximation of the rank function, which practically improves {\color{\MinEdt}recovery} limits on the rank {\color{\MinEdt}of the solution}. {\color{\MinEdt}This approximation leads to a non-convex program; thus, to avoid getting trapped in local solutions, we use the following scheme. Initially, a rough approximation of the rank function subject to the affine constraints is optimized. As the algorithm proceeds, finer approximations of the rank are optimized and the solver is initialized with the solution of the previous approximation until reaching the desired accuracy.}

On the theoretical side, benefiting from the spherical section property, we will show that the sequence of {\color{\MinEdt}the solutions of the} approximating function converges to the minimum rank solution. On the experimental side, it will be shown that {\color{\MinEdt}the proposed algorithm, termed SRF standing for Smoothed Rank Function,} can recover matrices which are unique solutions of the rank minimization problem and yet not recoverable by nuclear norm minimization. Furthermore, it will be demonstrated that, in completing partially observed matrices, the accuracy of SRF is considerably and consistently better than some famous algorithms {\color{\MinEdt}when the number of revealed entries is close to the minimum number of parameters that uniquely represent a low-rank matrix.}
\end{abstract}


\IEEEpeerreviewmaketitle

\section{Introduction}

\IEEEPARstart{T}{here} are many applications in signal processing and control theory which involve finding a matrix with minimum rank subject to linear constraints~\cite{RechFP10}. This task is usually referred to as the affine rank minimization (ARM) and includes \emph{Matrix Completion} (MC) as a special case. In the latter, we are interested in reconstructing a low-rank matrix from a subset of its entries. If the location of known entries follow certain random laws and the rank of the matrix is sufficiently small, one can uniquely recover the matrix with overwhelming probability~\cite{RechFP10, CandR09, RechXH11}.

One of the celebrated applications of affine rank minimization (or matrix completion) is \emph{Collaborative Filtering}~\cite{CandR09}. This technique is applied when a system tries to recommend goods to customers/users based on the available feedbacks of all the customers. In fact, the system learns the user preferences through the feedbacks and identifies similarities between them. As the number of factors affecting the user interests is much less than the total number of customers and products, the matrix whose $(i,j)${\color{\MinEdt}-th} entry represents the rating of the $i$-th user for the $j$-th product is expected to be low-rank. This could be efficiently used by the matrix completion techniques to predict the users' ratings for unrated items.

Applications of {\color{\FinEdt}affine} rank minimization in control theory include \emph{System Identification}~\cite{CandP10} and low-order realization of linear systems~\cite{FazeHB01}. In the former, the goal is to find an LTI system with minimum order that fits the available joint input-output observations of a multiple-input multiple-output system~\cite{LiuV09}.

In wireless sensor networks, due to limited energy resources and transmitting power, the sensors are able to communicate only with their neighboring sensors. These communications (e.g., received powers) determine the pairwise distances between sensors, which partially reveals the matrix of all pairwise distances. To localize the sensors in the network, one needs to estimate their distances from predefined anchor points which in turn requires completion of the distance matrix through the multi-dimensional scaling technique~\cite{CoxC94}. Interestingly, the rank of the pairwise distance matrix is small compared to its dimension~\cite{CandP10}.

Other areas to which affine rank minimization is applied include \emph{Machine Learning}~\cite{AmitFSU07}, \emph{Quantum State Tomography}~\cite{GrosLFBE10}, \emph{Spectrum Sensing}~\cite{MengYLHH11}, and \emph{Spatial Covariance} matrix completion~\cite{CandP10, ItoVOGS10}. The spatial covariance matrix is essential in estimating the directions of arrival of sources impinging on an array of antenna{\color{\MinEdt}s} using for example MUSIC~\cite{Schm86} or ESPRIT algorithms~\cite{RoyK89}.

The main difficulty of the {\color{\FinEdt}affine} rank minimization problem is due to the fact that the rank function is discontinuous and non-differentiable. Indeed, the optimization problem is NP-hard, and all available optimizers have doubly exponential complexity~\cite{ChistG84}. In~\cite{Faze02}, Fazel proposed to replace the rank of the matrix with its nuclear norm, which is defined as the sum of all singular values (SV). This modification is known to be the tightest convex relaxation of the rank minimization problem~\cite{CandP10} and can be implemented using a Semi Definite Program (SDP)~\cite{RechFP10}. Using similar techniques as in compressed sensing, it is recently shown that under mild conditions and with overwhelming probability, the  nuclear norm minimization (NNM) technique achieves the same solution as the original rank minimization approach~\cite{MohaFH11,OymaH10,KongTX11}.

Other approaches toward rank minimization consist of either alternative solvers instead of SDP in NNM or approximating the rank function using other forms rather than the nuclear norm. The FPCA method belongs to the first category and uses fixed point and Bergman iterative algorithm to solve NNM~\cite{MaGC11}. Among the examples of the second category, one can name {\color{\MinEdt} LMaFit \cite{WenYZ12}, BiG-AMP \cite{ParkSC13}, and OptSpace \cite{KeshMO10}.} It is also possible to generalize the greedy methods of compressive sensing to the rank minimization problem; for instance, ADMiRA~\cite{LeeB10} generalizes the CoSaMP~\cite{NeedT09}.

In this work, we introduce an iterative method that is based on approximating the rank function. However, in contrast to previous methods, the approximation is continuous and differentiable, is made finer at each iteration, and, asymptotically, will coincide with the rank function.
%{\color{\MinEdt}At this background,} it will be shown that this approach {\color{\HlCol}finds} %leads to finding
%solutions which are not obtainable by NNM, while they are unique rank minimizers.
Our method is inspired by the work of Mohimani et al~\cite{MohiBJ09} which uses smoothed $\ell_0$-norm\footnote{$\ell_0$-norm, not mathematically a vector norm, denotes the number of non-zero elements of a vector.} to obtain sparse solutions of underdetermined system of linear equations. {\color{\FinEdt}Nevertheless, }{\color{\RTwo_Col}the way SRF is extended from \cite{MohiBJ09}, and, particularly, the performance guarantees that are provided are among the contribution of our paper. Furthermore, in generalizing the method of \cite{MohiBJ09} to the ARM problem, we need to derive the gradient of the rank approximating functions in a closed form which is another novelty of the current work.}

A few preliminary results of this work have been presented in the conference paper~\cite{GhasMBJ11}. While~\cite{GhasMBJ11} was only devoted to the matrix completion problem, the current paper focuses on the more general problem of affine rank minimization. Furthermore, here, we present mathematical and experimental convergence analysis and consider more comprehensive numerical evaluation scenarios.

The reminder of this paper is organized as follows. In Section~\ref{sec:PrbFrm}, the ARM problem is formulated, and in Section~\ref{sec:algo}, the SRF algorithm is introduced. Section~\ref{sec:ThAna} is devoted to analyze the convergence properties of the SRF algorithm. In Section~\ref{sec:NumAna}, some experimental results of our algorithm are provided, and it will be compared empirically against {\color{\ROne_Col}some} well known algorithms. {\color{\MinEdt}Finally, Section \ref{sec:Con} concludes the paper.}

\section{Problem Formulation} \label{sec:PrbFrm}
The affine rank minimization problem generally is formulated as
\begin{equation}
\min_{\Xb}\rank(\Xb) \text{ subject to } \ACal(\Xb)=\bb \text{,} \label{RM}
\end{equation}
where $\Xb \in \Rbb^{n_1 \times n_2}$ is the decision variable, $\LOp$ is a known linear operator, and $\bb \in \Rbb^{m}$ is the observed measurement vector. {\color{\ROne_Col}The affine constraints $\ACal(\Xb)=\bb$ can be converted to
\begin{equation} \label{MatRep}
\Ab\vect(\Xb)=\bb,
\end{equation}
where $\mathbf{A} \in \Rbb^{m \times n_1 n_2}$ denotes the matrix representation of the linear operator $\ACal$ and $\vect(\Xb)$ denotes the vector in $\Rbb^{n_1 n_2}$ with the columns of $\Xb$ stacked on top of one another.}

The special case of matrix completion corresponds to the setting
\begin{equation}
\min_{\Xb}\rank(\Xb) \text{ subject to } [\Xb]_{ij} = [\mathbf{M}]_{ij} ~~ \forall(i,j) \in \Omega \text{,} \label{MC}
\end{equation}
where $\Xb$ is as in~\eqref{RM}, $\mathbf{M} \in \Rbb^{n_1 \times n_2}$ is the matrix whose entries are partially observed, $\Omega \subset \{1,2, ..., n_1\} \times \{1,2, ..., n_2\}$ is the set of the indexes of the observed entries of $\mathbf{M}$, and $[\Xb]_{ij}$ is the $(i,j)$-th entry of $\Xb$. Indeed, the constraints $[\Xb]_{ij} = [\mathbf{M}]_{ij}, \forall(i,j) \in \Omega,$ is an affine mapping which keeps some of the entries and discards others.

In the nuclear norm minimization, the rank function is replaced with the nuclear norm of the decision variable, leading to
\begin{equation}
\min_{\Xb} \|\Xb\|_* \text{ subject to } \ACal(\Xb)=\bb \text{,} \label{NM}
\end{equation}
where $\|\Xb\|_* \triangleq \sum_{i=1}^{r}{\sigma_i(\Xb)}$ is the nuclear norm, in which $r$ is the rank of the matrix $\Xb$, and $\sigma_i(\Xb)$ is the $i$-th largest singular value of the matrix $\Xb$. There is a strong parallelism between this rank minimization and $\ell_0$-norm minimization in compressive sensing~\cite{RechFP10}. In particular, minimizing the rank is equivalent to minimizing the number of non-zero singular values. Hence, \eqref{RM} can be reformulated as
\begin{equation}
\min_{\Xb} \|\SigbXb\|_0 \text{ subject to } \ACal(\Xb)=\bb \text{,}
\end{equation}
where $\SigbXb=(\sigma_1(\Xb),...,\sigma_n(\Xb))^T$ is the vector of all singular values, $\|\cdot\|_0$ denotes the $\ell_0$-norm, and $\nDef$.\footnote{Note that just $r$ entries of $\SigbXb$ are non-zero where $r$ is the rank of the matrix $\Xb$.}
Likewise, the nuclear norm is the $\ell_1$-norm of the singular value vector where the $\ell_1$-norm of a vector, denoted by $\|\cdot\|_1$, is the sum of the absolute values of its elements. This suggests the alternative form of
\begin{equation}
\min_{\Xb} \|\SigbXb\|_1 \text{ subject to } \ACal(\Xb)=\bb %\text{,}
\end{equation}
for \eqref{NM}. Based on this strong parallel, many results in compressive sensing theory (see for example~\cite{CandT05,CandRT06,Dono06,Zhan08}) have been adopted in the rank minimization problem~\cite{RechFP10,MohaFH11,OymaH10,OymaKH11}.

\section{The proposed algorithm} \label{sec:algo}
\subsection{The main idea} \label{sec:mainidea}
Our approach to solve the ARM problem is to approximate the rank with a continuous and differentiable function, and then to use a gradient descent algorithm to minimize it. The approximation is such that the error can be made arbitrarily small. {\color{\MinEdt}In contrast, note} that {\color{\FinEdt}the} nuclear norm is not differentiable~\cite{LewiS05} and its approximation error depends on the singular values of the matrix and cannot be controlled.

Instead of using a fixed approximation, we use a family $G_{\delta}: \Rbb^{n_1 \times n_2} \rightarrow \Rbb^{+}$ of approximations, where the index $\delta$ is a measure of approximation error and reflects the accuracy. The smaller $\delta$, the closer behavior of $G_{\delta}$ to the rank. For instance, $G_{0}$ stands for the errorless approximation; i.e., $G_{0}$ coincides with the rank function. We constrain the family to be continuous with respect to $\delta$. This helps in achieving the rank minimizer ($G_{0}$) by gradually decreasing $\delta$. Besides, to facilitate finding the minimizers of the relaxed problem, we require the $G_{\delta}$'s for $\delta>0$ to be differentiable with respect to the input matrix.

In order to introduce suitable $G_{\delta}$ families, we specify certain families of one-dimensional functions that approximate Kronecker delta function.

\newtheorem{Assump1}{Assumption}
\begin{Assump1} \label{Assump1}
Let  $f: \Rbb \rightarrow [0,1]$ and define $\fdel(x) = f(x/\delta)$ for all $\delta > 0$. The class $\{\fdel\}$ is said to satisfy the Assumption~\ref{Assump1}, if
\begin{enumerate}
\item[(a)] $f$ is real, symmetric unimodal, and analytic,

\item[(b)] $f(x) = 1 \Leftrightarrow x = 0$,

\item[(c)] $f''(0)<0$, and

\item[(d)] $\lim_{|x|\rightarrow \infty}f(x) = 0$.

\end{enumerate}
\end{Assump1}

It follows from Assumption~\ref{Assump1} that $x=0$ is the unique mode of all $\fdel$'s. This implies that $\fdel'(0) = 0$ for $\delta\neq 0$. In addition, $\{\fdel\}$ converge pointwise to Kronecker delta function as $\delta\rightarrow 0$, i.e.,
\begin{equation}\label{eq:fdelLim}
\lim_{\delta \to 0} \fdel(x)= \left\{
	\begin{array}{rl}
	0 & \text{if } x \neq 0,\\
	1 & \text{if } x = 0.
	\end{array} \right.
\end{equation}

The class of Gaussian functions, which is of special interest in this paper, is defined as
\begin{equation} \label{GaussianDef}
\fdel(x)=\exp(-\frac{x^2}{2\delta^2}).
\end{equation}
It is not difficult to verify the constraints of Assumption~\ref{Assump1} for this class. Other examples include $\fdel(x)=1-\tanh(\frac{x^2}{2\delta^2})$ and $\fdel(x)=\frac{\delta^2}{x^2+\delta^2}$.

To extend the domain of $\{\fdel\}$ to matrices, let define
\begin{equation} \label{FDef}
\Fdel(\Xb) = h_{\delta}\big(\SigbXb\big)=\sumn\fdel\big(\sigma_i(\Xb)\big),
\end{equation}
where $\nDef$ and $h_{\delta}:\Rbb^{n} \rightarrow \Rbb$ is defined as $h_{\delta}(\xb)= \sumn f_{\delta}(x_i)$. Since $\fdel$ is an approximate Kronecker delta function, $\Fdel(\Xb)$ yields an estimate of the number of zero singular values of $\Xb$. Consequently, it can be concluded that $\rank(\Xb) \approx n - \Fdel(\Xb)$, and the ARM problem can be relaxed to
\begin{equation} \label{Gmin}
\min_{\Xb}\big(G_{\delta}(\Xb) = n-\Fdel(\Xb) \big) \text{ subject to } \ACal(\Xb)=\bb \text{,}
\end{equation}
or equivalently
\begin{equation} \label{Fmax}
\max_{\Xb}\Fdel(\Xb) \text{ subject to } \ACal(\Xb)=\bb \text{.}
\end{equation}

The advantage of maximizing $\Fdel$ compared to minimizing the rank is that $\Fdel$ is smooth and we can apply gradient methods. However, for small values of $\delta$ where $G_{\delta}$ is a relatively good approximate of the rank function, $\Fdel$ has many local maxima, which are likely to trap gradient methods.

To avoid local maxima\footnote{{\color{\RFor_Col}For any finite $\delta > 0$, $\Fdel(\cdot)$ is not a concave function, and, throughout the paper, a local maximum of $\Fdel(\cdot)$ denotes a point which is locally and not, at the same time, globally maximum.}}, we initially apply a large $\delta$. Indeed, we will show in Theorem~\ref{InitThm} that under, Assumption~\ref{Assump1}, $\Fdel$ becomes {\color{\MinEdt}concave} as $\delta \to \infty$ and \eqref{Fmax} will have a unique solution. Then we gradually decrease $\delta$ to improve the accuracy of approximation. For each new value of $\delta$, we initialize the maximization of $\Fdel$ with the result of \eqref{Fmax} for the previous value of $\delta$. From the continuity of $\{\fdel\}$ with respect to $\delta$, it is expected that the solutions of \eqref{Fmax} for $\delta_{i}$ and $\delta_{i+1}$ are close, when $\delta_{i}$ and $\delta_{i+1}$ are close. In this fashion, the chance of finding a local maximum instead of a global one is decreased.  This approach for optimizing non-convex functions is known as Graduated Non-Convexity (GNC)~\cite{BlakZ87}, and was used in~\cite{MohiBJ09} to minimize functions approximating the $\ell_0$-norm.

\subsection{Gradient Projection}
For each $\delta$ in the decreasing sequence, to maximize $\Fdel$ with equality constraints, we use the Gradient Projection (GP) technique~\cite{Bert99}. In GP, the search path at each iteration is obtained by projecting back the ascent (or descent) direction onto the feasible set~\cite{Bert99}. In other words, at each iteration, one has $\Xb \leftarrow \mathcal{P}\big(\Xb + \mu_j\nabla\Fdel(\Xb)\big)$, where $\mathcal{P}$ denotes the orthogonal projection onto the affine set defined by linear constraints $\ACal(\Xb)=\bb$, and $\mu_j$ is the step{\color{\FinEdt}--}size of the $j$-th iteration. As the feasible set is affine, several methods can be exploited to implement the projection $\mathcal{P}$. For example, one can store the QR factorization of the matrix implementation of $\ACal$ for fast implementation of the back projection, or, alternatively, a least-squares problem can be solved at each step~\cite{RechFP10}. The closed form solution of the least-squares problem can be found in Appendix~\ref{appA}.

To complete the GP step, we should derive the gradient of the approximating functions with respect to the matrix $\Xb$. Surprisingly, although $\sigma_i(\Xb), i = 1, ..., n$ and $\|\Xb\|_*$ are not differentiable functions of $\Xb$~\cite{LewiS05}, the following theorem shows that one can find functions $\Fdel=h_{\delta} \circ \SigbXb$ which are differentiable under the absolutely symmetricity of the $h_{\delta}$. Before stating the theorem, recall that a function $h:\Rbb^q \rightarrow [-\infty,+\infty]$ is called \emph{absolutely symmetric}~\cite{Lewi95} if $h(\mathbf{x})$ is invariant under arbitrary permutations and sign changes of the components of $\mathbf{x}$.

\newtheorem{Thm1}{Theorem}
\begin{Thm1} \label{GradThm}
Suppose that $F : \Rbb^{n_1 \times n_2} \rightarrow \Rbb$ is represented as $F(\Xb)=h\big(\SigbXb\big)=h \circ \SigbXb$, where $\Xb \in \Rbb^{n_1 \times n_2}$ with the Singular Value Decomposition (SVD) $\Xb = \mathbf{U} \text{diag}(\sigma_1,...,\sigma_n)\mathbf{V}^T$, $\SigbXb:\Rbb^{n_1 \times n_2} \rightarrow \Rbb^n$ has the SVs of the matrix $\Xb$, $\nDef$, and $h:\Rbb^{n} \rightarrow \Rbb$ is absolutely symmetric {\color{\FinEdt}and differentiable}. Then the gradient of $F(\Xb)$ at $\Xb$ is
\begin{equation} \label{GradDef}
\frac{\partial{F(\Xb)}}{\partial{\Xb}}=\mathbf{U} \text{diag}(\boldsymbol{\theta}) \mathbf{V}^T,
\end{equation}
where $\boldsymbol{\theta} = \frac{\partial h(\yb)}{\partial \yb}\lvert_{\yb=\SigbXb}$ denotes the gradient of $h$ at $\SigbXb$.

\vspace{0.2em}
\hspace{0.6em}
\emph{Informal Proof}: In~\cite[Cor. 2.5]{Lewi95}, it is shown that if a function $h$ is absolutely symmetric and the matrix $\Xb$ has $\SigbXb$ in the domain of $h$, then the subdifferential\footnote{To see the definition of subdifferential and subgradient of non-convex functions, refer to ~\cite[Sec. 3]{Lewi96}.}  of $F$ is given by
\begin{equation}
\partial \big(h \circ \SigbXb \big) = \{ \mathbf{U} \text{diag}(\boldsymbol{\theta}) \mathbf{V}^T | \boldsymbol{\theta} \in \partial h\big(\SigbXb\big)\}.
\end{equation}
Since $h$ is differentiable, $\partial h\big(\SigbXb\big)$ is a singleton and consequently $\partial \big(h \circ \SigbXb \big)$ becomes a singleton. When the subdifferential of a non-convex function becomes singleton, the function is intuitively expected to be differentiable with the subgradient as its gradient.\footnote{For a convex function, the subdifferential is singleton iff the function is differentiable~\cite{Rock70}.} Nevertheless, to the best of our knowledge, there is no formal proof.
Provided that this intuition is true, then $\partial \big(h \circ \SigbXb \big)$ will be converted to $\nabla \big(h \circ \SigbXb \big)$ and equation~\eqref{GradDef} is obtained.
\begin{proof}[Formal Proof] Equation~\eqref{GradDef} can be obtained directly from the ``if part" of~\cite[Thm. 3.1]{Lewi95}, which does not require convexity of $h$ as stated in its proof.
\end{proof}
\end{Thm1}

\newtheorem{Cor1}{Corollary}
\begin{Cor1}
For the Gaussian function family given in~\eqref{GaussianDef}, the gradient of $\Fdel(\Xb)$ at $\Xb$ is
\begin{equation} \label{GGradDef}
\frac{\partial{\Fdel(\Xb)}}{\partial{\Xb}}=\mathbf{U} \text{diag}(-\frac{\sigma_1}{\delta^2}\text{e}^{-\sigma_1^2/2\delta^2},...,-\frac{\sigma_n}{\delta^2}\text{e}^{-\sigma_n^2/2\delta^2}) \mathbf{V}^T.
\end{equation}
\begin{proof}
$f_{\delta}$ is an even function for the Gaussian family; therefore, $h_{\delta}$ becomes an absolutely symmetric function. As a result, Theorem~\ref{GradThm} proves~\eqref{GGradDef}.
\end{proof}
\end{Cor1}

\subsection{Initialization}
Naturally, we initialize the GNC procedure by the solution of~\eqref{Fmax} corresponding to $\delta \to \infty$. This solution can be found from the following theorem.

\newtheorem{Thm2}[Thm1]{Theorem}
\begin{Thm2} \label{InitThm}
Consider a class of one variable functions $\{\fdel\}$ satisfying the Assumption~\ref{Assump1}. For the rank approximation problem~\eqref{Fmax}, let $\XbF = \argmin \{\|\Xb\|_F \;|\; \ACal(\Xb)=\bb\}$, then
\begin{equation}
\lim_{\delta \rightarrow \infty} \argmax \{\Fdel(\Xb) \;|\; \ACal(\Xb)=\bb\} =\XbF,
\end{equation}
where $\|\cdot\|_F$ denotes the matrix Frobenius norm.
\end{Thm2}

There is a simple interpretation of the solution of~\eqref{Fmax} for the Gaussian family when $\delta$ approaches $\infty$. As $e^{-x} \approx 1-x$ for small values of $x$,
\begin{equation*}
\Fdel(\Xb)=\sumn e^{-\sigma_i^2(\Xb)/2\delta^2}\approx n - \sumn \sigma_i^2(\Xb)/2\delta^2
\end{equation*}
for $\delta \gg \sigma_i(\Xb)$. Consequently,
\begin{IEEEeqnarray*}{rCl}
\IEEEeqnarraymulticol{3}{l}{ \argmax \{\Fdel(\Xb) \;|\; \ACal(\Xb)=\bb\} \approx }\\
& & \argmin \{\sumn \sigma_i^2(\Xb) | \ACal(\Xb)\!=\!\bb\}\! =\!\argmin \{\|\Xb\|_F | \ACal(\Xb)\!=\!\bb\}.
\end{IEEEeqnarray*}
The proof is left to Appendix~\ref{appB}.

The following corollary is an immediate result of the above theorem.
\newtheorem{Cor2}[Cor1]{Corollary}
\begin{Cor2}
For the matrix completion problem, the initial solution of the SRF algorithm is $\XbF$ with the following definition:
\begin{equation}
[\XbF]_{ij}=\left\{
\begin{array}{cc}
 [\mathbf{M}]_{ij} & (i,j) \in \Omega,\\
 0 & (i,j)  \notin \Omega.
\end{array} \right.,
\end{equation}
where $\mathbf{M}$ and $\Omega$ are as defined in~\eqref{MC}.
\end{Cor2}

\subsection{The Final Algorithm}

\newif\ifonecol
\onecolfalse
\begin{figure}
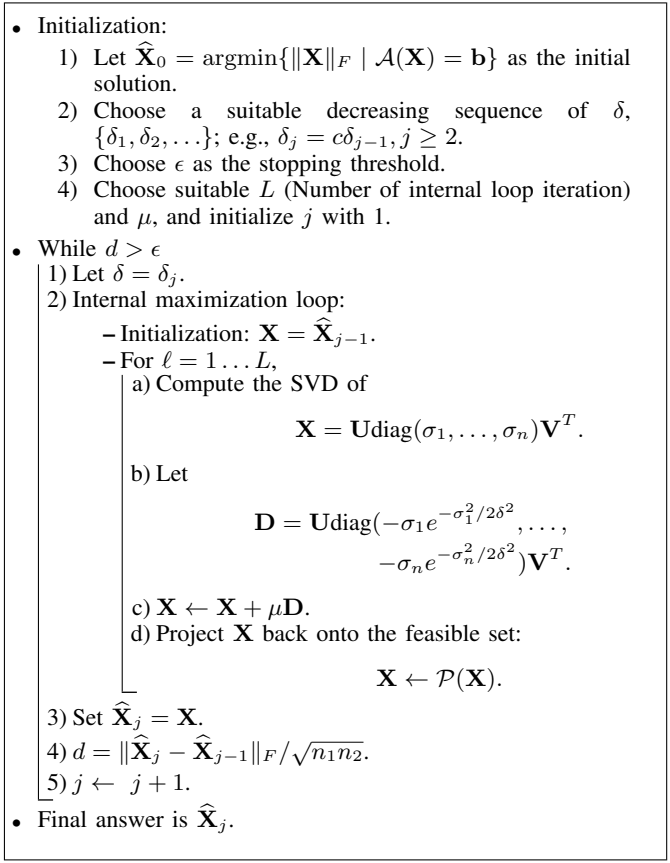

  \centering \vrule%
  \ifonecol
    \begin{minipage}{9.7cm} % For 1 column paper (peerreview)
  \else
    \begin{minipage}{8.7cm} % For 2 column paper (peerreview)
  \fi
\hrule \vspace{0.5em}%\centering
\hspace*{-1.1em}
  \ifonecol
    \begin{minipage}{9.5cm} % For 1 column paper (peerreview)
  \else
    \begin{minipage}{8.5cm} % For 2 column paper (peerreview)
  \fi
      {
        \small
        \def\baselinestretch{1}

        \begin{itemize}
        \item Initialization:

           \begin{enumerate}
             \item Let $\Xbh_0=\argmin \{\|\Xb\|_F \;|\; \ACal(\Xb)=\bb\}$ as the initial solution.
             \item Choose a suitable decreasing sequence of $\delta$, $\{\delta_1, \delta_2,\ldots\}$; e.g., $\delta_j=c\delta_{j-1}, j \geq 2$.
             \item Choose $\epsilon$ as the stopping threshold.
             \item Choose suitable $L$ (Number of internal loop iteration) and $\mu$, and initialize $j$ with 1.
           \end{enumerate}

        \item While $d > \epsilon$ \vspace{1mm}\\
        \vrule\hspace{-1em}
        \begin{minipage}{8.2cm}
          \begin{enumerate}
             \item \hspace{-0.8em} Let $\delta=\delta_j$.
             \item \hspace{-0.8em} Internal maximization loop:
             \begin{itemize}
               \item \hspace{-0.8em} Initialization: $\Xb=\Xbh_{j-1}$.
               \item \hspace{-0.8em} For $\ell=1\dots L$, \vspace{0.5mm} \\
                  \hspace*{-0.4em}\vrule\hspace{-1em}
                  \begin{minipage}{8cm}
                    \begin{enumerate}
					   \item \hspace{-0.8em} Compute the SVD of
					   \begin{equation*}
	   					   \Xb=\mathbf{U}\text{diag}(\sigma_1,\ldots,\sigma_n)\mathbf{V}^T.
					   \end{equation*}
                       \item \hspace{-0.8em} Let
                       \begin{eqnarray*}
                       \mathbf{D}=\mathbf{U}\text{diag}(-\sigma_1e^{-\sigma_1^2/2\delta^2},\ldots, \\
                       -\sigma_n e^{-\sigma_n^2/2\delta^2})\mathbf{V}^T.
                       \end{eqnarray*}
                       \item \hspace{-0.8em} $\Xb\leftarrow \Xb+\mu \mathbf{D}$.
                       \item \hspace{-0.8em} Project $\Xb$ back onto the feasible set:
                          \begin{equation*}
                             \Xb\leftarrow\mathcal{P}(\Xb).
                          \end{equation*} %\vspace*{.05em}
                    \end{enumerate}
                  \end{minipage}\\\hspace*{-0.4em}\rule{2mm}{.5pt}
             \end{itemize}
             \item \hspace{-0.8em} Set $\Xbh_{j}=\Xb$.\vspace{0.2em}
             \item \hspace{-0.8em} $d=\|\Xbh_j-\Xbh_{j-1}\|_F/\sqrt{n_1n_2}$.\vspace{0.2em}
             \item \hspace{-0.8em} $j \leftarrow\ j + 1$.\vspace{0.2em}
          \end{enumerate}
        \end{minipage}\\\rule{2mm}{.5pt}
        \item Final answer is $\Xbh_j$.
        \end{itemize}
      }
    \end{minipage}
    \vspace{1em} \hrule
  \end{minipage}\vrule \\
\caption{The SRF Algorithm.} \label{fig:algorithm}
\end{figure}

The final algorithm is obtained by applying the main idea, initial solution, and gradient projection to the Gaussian function given in~\eqref{GaussianDef}. Fig.~\ref{fig:algorithm} depicts the algorithm. In the sequel, we briefly review some remarks about the parameters used in the implementation of the algorithm. Most of these remarks correspond to similar remarks for the SL0 algorithm~\cite{MohiBJ09} and are presented here for the sake of completeness.

\emph{Remark 1.} It is not necessary to wait for the convergence of the internal steepest ascent loop because as explained in Section~\ref{sec:mainidea} for each value of $\delta$, it is just needed to get close to the global maximizer of $\Fdel$ to avoid local maxima. Therefore, the internal loop is only repeated for a fixed number of times ($L$).

\emph{Remark 2.} After initiating the algorithm with the minimum Frobenius norm {\color{\MinEdt}solution}, the first value of $\delta$ may be set to about two to four times of the largest SV of $\Xbh_0$ (the initial guess). If we take $\delta > 4\max_i\big(\sigma_i(\Xbh_0)\big)$, then $\exp\big(-\sigma_i^2(\Xbh_0)/2\delta^2\big) > 0.96 \approx 1$ for $1 \leq i \leq n$. Thus, this $\delta$ value acts virtually like $\infty$ for all SVs of $\Xbh_0$. {\color{\MinEdt}In addition}, the decreasing sequence can be adjusted to $\delta_j=c\delta_{j-1}, j \geq 2$, where $c$ generally is chosen between 0.5 and 1.

\emph{Remark 3.} This remark is devoted to the selection of $\mu_j$, step--size parameter. Typically, in a gradient ascent algorithm, $\mu_j$ should be chosen small enough to follow the ascent direction. Furthermore, reducing $\delta$ results in more fluctuating behaviour of the rank approximating function. Therefore, to avoid large steps which cause jumps over the maximizer, one should choose smaller value{\color{\FinEdt}s} of step--size for smaller values of $\delta$. Following the same reasoning as in~\cite[Remark 2]{MohiBJ09}, a good choice is to decrease $\mu_j$ proportional to $\delta^2$; that is, $\mu_j=\mu\delta^2$, where $\mu$ is a constant. By letting $\mu_j = \mu \delta^2 $, the gradient step can be reduced to
\begin{equation*}
\Xb \leftarrow \Xb - \mu \mathbf{U} \text{diag}(\sigma_1 e^{-\sigma_1^2/2\delta^2}, \ldots, \sigma_n e^{-\sigma_n^2/2\delta^2})\mathbf{V}^T.
\end{equation*}

\emph{Remark 4.} The distance between the solutions at the two consecutive iterations is the criterion to stop the algorithm. That is, if $d \triangleq \|\Xbh_j - \Xbh_{j-1}\|_F / \sqrt{n_1 n_2}$ is smaller than some tolerance ($\epsilon$), the iterations are ended and $\Xbh_j$ becomes the final solution.

\section{Convergence Analysis} \label{sec:ThAna}
{\color{\RTwo_Col} Noting that the original problem is NP-hard and we are dealing with maximizing non-concave functions, a complete and thorough convergence analysis would be beyond the scope of this paper. We believe that similar to \cite{MohiBGJ10} which examines the global convergence properties of the SL0 algorithm \cite{MohiBJ09}, it would be possible to analyze the convergence of the SRF algorithm to the global solution. However, in this paper, we only study a simplified convergence analysis, and the complete analysis is left for a future work.

For the simplified analysis,} in the sequel, it is assumed that the internal loop has been converged to the global maximum, and we prove that {\color{\MinEdt}this global} solution converges to the minimum rank solution as $\delta$ goes to zero. {\color{\RTwo_Col}This analysis helps us to characterize the conditions under which
\begin{equation} \label{argmax1}
\lim_{\delta \to 0} \argmax \{\Fdel(\Xb) \;|\; \ACal(\Xb) = \bb\}
\end{equation}
is equivalent to
\begin{equation} \label{argmax2}
\argmax \{\lim_{\delta \to 0} \Fdel(\Xb)\; |\; \ACal(\Xb) = \bb\}.
\end{equation}
The equivalence of \eqref{argmax1} and \eqref{argmax2} is of particular importance since it shows that the idea of SRF corresponding to optimization of \eqref{argmax1} is indeed the case and leads to finding the solution of program \eqref{argmax2} which is identical to the original affine rank minimization problem defined in \eqref{RM}.}

The following results and proofs are not direct extension of the convergence results of~\cite{MohiBJ09} and are more tricky to obtain, though our exposition follows the same line of presentation.

We start the convergence analysis by the definition of the Spherical Section Property (SSP), used in the analysis of uniqueness of the rank and nuclear norm minimization~\cite{MohaFH11}, and a lemma which makes this abstract definition clearer.

\newtheorem{Def1}{Definition}
\begin{Def1} \label{Def1}
\textbf{Spherical Section Property}~\cite{MohaFH11,DvijF10}. The spherical section constant of a linear operator $\LOp$ is defined as
\begin{equation}
\Delta(\ACal)=\min_{\mathbf{Z} \in \nullS(\ACal) \setminus \{\mathbf{0}\}} \frac{\|\mathbf{Z}\|_*^2}{\|\mathbf{Z}\|_F^2}.
\end{equation}
Further, $\ACal$ is said to have the $\Delta$-spherical section property if $\Delta(\ACal) \geq \Delta$.
\end{Def1}

{\color{\ROne_Col}Definition \ref{Def1} extends a similar concept in the compressive sensing framework where it is shown that many randomly generated sensing matrices possesses the SSP with high probability \cite{Zhan08}. Although extending a similar theoretical result to the matrix case is a topic of interest, \cite{DvijF10} proves that if all entries of the matrix representation of $\ACal$ are identically and independently distributed from a zero-mean, unit-variance Gaussian distribution, then, under some mild conditions, $\ACal$ possesses {\color{\FinEdt}the} $\Delta$-spherical section property with overwhelming probability.}

\newtheorem{Lemma1}{Lemma}
\begin{Lemma1}
Assume $\ACal$ has the $\Delta$-spherical section property. Then, for any $\Xb \in \nullS(\ACal)\setminus \{\mathbf{0}\}$, we have $\rank(\Xb) \geq \Delta$.
\begin{proof}
Since $\Xb$ belongs to $\nullS(\ACal)$, one can write
\begin{equation*}
\frac{\|\Xb\|_{*}}{\|\Xb\|_F} \geq \sqrt{\Delta} \Rightarrow \|\Xb\|_* \geq \sqrt{\Delta} \|\Xb\|_F.
\end{equation*}
It is also known that $\sqrt{\rank(\Xb)} \|\Xb\|_F \geq \|\Xb\|_*$, see for example~\cite{HornJ85}. Putting them together, we have
\begin{equation*}
\|\Xb\|_{*} \geq \sqrt{\Delta}\frac{\|\Xb\|_*}{\sqrt{\rank(\Xb)}} \Rightarrow \rank(\Xb) \geq \Delta
\end{equation*}
or $\rank(\Xb) \geq \lceil\Delta\rceil$, where $\lceil\Delta\rceil$ denotes the smallest integer greater than or equal to $\Delta$.
\end{proof}
\end{Lemma1}

The above lemma shows that if $\Delta$ is large, the null space of $\ACal$ does not include low-rank matrices. Such subspaces are also known as almost Euclidean subspaces~\cite{Zhan08}, in which the ratio of $\ell_1$-norm to $\ell_2$-norm of elements cannot be small.

\newtheorem{Thm3}[Thm1]{Theorem}
\begin{Thm3}[\cite{DvijF10}] \label{RMUniq}
Suppose $\ACal$ has the $\Delta$-spherical property, and $\Xb_0 \in \mathbb{R}^{n_1 \times n_2}$ satisfies $\ACal(\Xb_0)=\bb$. If $\rank(\Xb_0)<\frac{\Delta}{2}$, then $\Xb_0$ is the unique solution of problem~\eqref{RM}.
\end{Thm3}

\newtheorem{Lemma2}[Lemma1]{Lemma}
\begin{Lemma2} \label{Lemma2}
Assume $\LOp$ has {\color{\FinEdt}the} $\Delta$-spherical section property, and set $\nDef$. Let $\Xb$ be any element in $\nullS(\ACal)$ and $\sigma_1, ..., \sigma_n$ represent its singular values. Then, for any subset $\ICal$ of $\{1,...,n\}$ such that $|\ICal|+\Delta > n$,
\begin{equation}
\frac{\sum_{i\in\ICal} \sigma_i}{(\sumn \sigma_i^2)^{0.5}} \geq \sqrt{\Delta}-\sqrt{n-|\ICal|},
\end{equation}
where $|\cdot|$ denotes the cardinality of a set.
\begin{proof}
If $\ICal=\{1,...,n\}$, then it is clear that $\frac{\sumn \sigma_i}{(\sumn \sigma_i^2)^{0.5}} \geq \sqrt{\Delta}$, since the $\Delta$-spherical section property holds. Otherwise, if $|\ICal| < n$, the $\Delta$-spherical section property implies that
\begin{equation*}
\sqrt{\Delta} \leq \frac{\|\Xb\|_*}{\|\Xb\|_F}=\frac{\sumn \sigma_i}{(\sumn \sigma_i^2)^{0.5}}.
\end{equation*}
For the sake of simplicity, let us define
\begin{equation*}
\alpha_i=\frac{\sigma_i}{(\sumn \sigma_i^2)^{0.5}}.
\end{equation*}
This shows that
\begin{equation*}
1=\sumn\alpha_i^2 \geq \sum_{i \notin \ICal} \alpha_i^2 \geq \frac{(\sum_{i \notin \ICal}\alpha_i)^2}{n-|\ICal|},
\end{equation*}
where we used the inequality $\forall \mathbf{z} \in \Rbb^p , \|\mathbf{z}\|_{1}^{2} \leq p\|\mathbf{z}\|_{2}^{2}$ {\color{\RTre_Col}\cite{HornJ85}}. Hence, it can be concluded that
\begin{equation*}
\sum_{i \notin \ICal} \alpha_i \leq \sqrt{n-|\ICal|}.
\end{equation*}
On the other hand, it is known that
\begin{equation*}
\sqrt{\Delta} \leq \sum_{i \in \ICal} \alpha_i + \sum_{i \notin \ICal} \alpha_i \leq \sum_{i \in \ICal} \alpha_i + \sqrt{n-|\ICal|},
\end{equation*}
which confirms that
\begin{equation*}
\frac{\sum_{i\in\ICal} \sigma_i}{(\sumn \sigma_i^2)^{0.5}} = \sum_{i \in \ICal} \alpha_i \geq \sqrt{\Delta} - \sqrt{n-|\ICal|}.
\end{equation*}
\end{proof}
\end{Lemma2}

\newtheorem{Cor3}[Cor1]{Corollary}
\begin{Cor3} \label{Cor3}
If $\LOp$ has {\color{\FinEdt}the} $\Delta$-spherical section property, $\nDef$, and $\Xb \in \nullS(\ACal)$ has at most $\UDelmOne$ singular values greater than $\alpha$, then
\begin{equation*}
\|\Xb\|_F \leq \frac{n\alpha}{\sqrt{\Delta} - \sqrt{\UDelmOne}}.
\end{equation*}
\begin{proof}
At least $n- \UDelmOne$ singular values of $\Xb$ are less than or equal to $\alpha$. If $\ICal$ denotes the indices of singular values not greater than $\alpha$, then by using Lemma~\ref{Lemma2}, we will have
\begin{eqnarray*}
\frac{\sum_{i\in\ICal} \sigma_i}{(\sumn \sigma_i^2)^{0.5}} \geq \sqrt{\Delta} - \sqrt{n-n+\UDelmOne} \Rightarrow \\
\|\Xb\|_F(\sqrt{\Delta}-\sqrt{\UDelmOne}) \leq \sum_{i \in \ICal} \sigma_i  \leq n\alpha,
\end{eqnarray*}
which proves that
\begin{equation*}
\|\Xb\|_F \leq \frac{n\alpha}{\sqrt{\Delta} - \sqrt{\UDelmOne}}.
\end{equation*}
\end{proof}
\end{Cor3}

\newtheorem{Lemma3}[Lemma1]{Lemma}
\begin{Lemma3} \label{Lemma3}
Assume $\LOp$ has {\color{\FinEdt}the} $\Delta$-spherical section property, $f_{\delta}(\cdot)$ is a member of the class that satisfies Assumption~\ref{Assump1}, and define $F_{\delta}$ as in \eqref{FDef} and $\nDef$. Let $\mathcal{X}=\{\Xb|\ACal(\Xb)=\bb\}$ contain a solution $\Xb_0$ with $\rank(\Xb_0) = r_0 < \frac{\Delta}{2}$. Then, for any $\widehat{\Xb} \in \mathcal{X}$ that satisfies
\begin{equation} \label{FGCond}
F_{\delta}(\widehat{\Xb}) \geq n - \left(\UDelmOne-r_0\right),
\end{equation}
we have that
\begin{equation*}
\|\Xb_0 - \widehat{\Xb}\|_F \leq \frac{n\alpha_{\delta}}{\sqrt{\Delta}-\sqrt{\UDelmOne}},
\end{equation*}
where $\alpha_{\delta}=\big|\fdel^{-1}(\frac{1}{n})\big|$.
\begin{proof}
First, note that due to Assumption~\ref{Assump1}, $\fdel(x)$ takes all the values in $]0,1[$ exactly twice; once with a positive $x$ and once with a negative one. Because of the symmetry, the two have the same modulus; therefore, $\alpha_{\delta}$ is well-defined.

Let us denote the singular values of $\Xb_0$ and $\widehat{\Xb}$ by $\sigma_1\geq \dots\geq \sigma_n$ and $\hat{\sigma}_1\geq \dots\geq \hat{\sigma}_n$, respectively. Define $\ICal_{\alpha}$ as the set of indices $i$ for which $\hat{\sigma}_i > \alpha_{{\color{\RTwo_Col}\delta}}$. Now, we have that
\begin{align*}
F_{\delta}(\widehat{\Xb})& = \sumn \fdel(\hat{\sigma}_i) \\
& = \underbrace{\sum_{i \in \ICal_{\alpha}} \underbrace{f_{\delta}(\hat{\sigma}_i)}_{ < \frac{1}{n}}}_{ < n \frac{1}{n}=1 } +
   \underbrace{\sum_{i \notin \ICal_{\alpha}} \underbrace{f_{\delta}(\hat{\sigma}_i)}_{ \leq 1}}_{ \leq n - |\ICal_{\alpha}| } \\
& < n - |\ICal_{\alpha}| + 1 \nonumber.
\end{align*}
On the other hand, $F_{\delta}(\widehat{\Xb}) \geq n - \left( \UDelmOne - r_0 \right )$; therefore,
\begin{IEEEeqnarray*}{rCl}
n - \left( \UDelmOne - r_0 \right ) & < & n - |\ICal_{\alpha}| + 1 \\
\Rightarrow |\ICal_{\alpha}| & < & \left( \UDelmOne - r_0 \right ) + 1\\
\Rightarrow |\ICal_{\alpha}|  & \leq & \UDelmOne - r_0.
\end{IEEEeqnarray*}
This means that at most $\UDelmOne - r_0$ singular values of $\widehat{\Xb}$ are greater than $\alpha_{\delta}$.
Define
\begin{align*}
\mathbf{H}_{0}=\left[\begin{array}{ll}
\mathbf{0} & \Xb_0 \\
\Xb_0^{T} & \mathbf{0}
\end{array}\right], ~~~ \widehat{\mathbf{H}}=\left[\begin{array}{ll}
\mathbf{0} & \widehat{\Xb} \\
\widehat{\Xb}^{T} & \mathbf{0}
\end{array}\right].
\end{align*}
In fact, $\mathbf{H}_{0}$ and $\widehat{\mathbf{H}}$ are {\color{\MinEdt}symmetric} matrices that contain the singular values of $\Xb_{0}$ and $\Xbh$, respectively, as their $n$ largest eigenvalues and their negatives as the $n$ smallest eigenvalues {\color{\MinEdt}\cite{HornJ85}}. Next, we apply Weyl's eigenvalue inequality {\color{\MinEdt}\cite{HornJ85}} as
\begin{align*}
\lambda_{\UDelmOne +1}(\mathbf{H}_{0} -\widehat{\mathbf{H}} ) &\leq \lambda_{r_0+1}(\mathbf{H}_{0}) +\lambda_{\UDelmOne - r_0 + 1}(-\widehat{\mathbf{H}} ) \nonumber\\
&= \sigma_{r_0+1} +\hat{\sigma}_{\UDelmOne - r_0 + 1}\nonumber\\
&= \hat{\sigma}_{\UDelmOne - r_0 + 1} \leq \alpha_{\delta},
\end{align*}
where $\lambda_{i}(\cdot)$ stands for the $i$-th largest eigenvalue. This reveals the fact that $(\Xb_0 - \widehat{\Xb})$ has at most $\UDelmOne$ singular values greater than $\alpha_{\delta}$. Since $(\Xb_0 - \widehat{\Xb})$ is in the null space of $\ACal$, Corollary~\ref{Cor3} implies that
\begin{equation*}
\|\Xb_0 - \widehat{\Xb}\|_F \leq \frac{n\alpha_{\delta}}{\sqrt{\Delta}-\sqrt{\UDelmOne}}.
\end{equation*}
\end{proof}
\end{Lemma3}

\newtheorem{Cor4}[Cor1]{Corollary}
\begin{Cor4} \label{AlphaG}
For the Gaussian function family given in~\eqref{GaussianDef}, if~\eqref{FGCond} holds for a solution $\Xbh \in \mathcal{X}$, then
\begin{equation*}
\|\Xbh - \Xb_0\|_F \leq \frac{n \delta \sqrt{2 \ln n}}{\sqrt{\Delta}-\sqrt{\UDelmOne}}.
\end{equation*}
\end{Cor4}

\newtheorem{Lemma4}[Lemma1]{Lemma}
\begin{Lemma4} \label{Lemma4}
Let $\fdel, \Fdel, \mathcal{X}, \text{and}, \Xb_0$ be as defined in Lemma~\ref{Lemma3} and assume $\Xb_{\delta}$ be the maximizer of $\Fdel(\Xb)$ on $\mathcal{X}$. Then $\Xb_{\delta}$ satisfies~\eqref{FGCond}.
\begin{proof}
One can write that
\begin{align*}
\Fdel(\Xb_{\delta}) & \geq \Fdel(\Xb_0) \\
& \geq n - r_0 \\
& \geq n - \left( \UDelmOne-r_0 \right).
\end{align*}
The first inequality comes from the fact that $\Xb_{\delta}$ is the maximizer of the $\Fdel(\Xb)$, and the second one is true because $\Xb_0$ has $(n-r_0)$ singular values equal to zero; thus, in the summation $F_{\delta}(\Xb)=\sumn f_{\delta}(\sigma_i)$, there are $(n-r_0)$ ones. Hence, $\Fdel(\Xb_0) \geq n - r_0$. To see the last inequality, note that $2r_0 < \Delta$ and $\Delta \leq \UDelmOne  + 1$.
Thus, it can be concluded that $2r_0 < \UDelmOne + 1$ which results in $2r_0 \leq \UDelmOne$ because $r_0 \in \mathbb{N}$. Finally, $r_0 \leq \UDelmOne - r_0$ which implies that $n - (\UDelmOne - r_0) \leq n - r_0$.
\end{proof}
\end{Lemma4}

Lemma~\ref{Lemma4} and Corollary~\ref{AlphaG} together prove that for the Gaussian family,
\begin{equation*}
\lim_{\delta \to 0} \argmax \{\Fdel(\Xb) \;|\; \ACal(\Xb)=\bb\}=\Xb_0.
\end{equation*}
In Theorem~\ref{theo4}, we extend this result to all function classes that satisfy Assumption \ref{Assump1}.

\newtheorem{Thm4}[Thm1]{Theorem}
\begin{Thm4}\label{theo4}
Suppose $\LOp$ has {\color{\FinEdt}the} $\Delta$-spherical property and $\{\fdel\}$ satisfies Assumption~\ref{Assump1}, and define $\mathcal{X}, \Fdel, \text{ and } \Xb_0$ as in Lemma~\ref{Lemma3}. If $\Xb_{\delta}$ represents the maximizer of $\Fdel(\Xb)$ over $\mathcal{X}$, then
\begin{equation*}
\lim_{\delta \to 0} \Xb_{\delta} = \Xb_{0}.
\end{equation*}

\begin{proof}
By combining Lemma~\ref{Lemma3} and Lemma~\ref{Lemma4}, we obtain that
\begin{equation}\label{eq:ErrorBound}
\|\Xb_0 - \Xb_{\delta}\|_F \leq \frac{n\alpha_{\delta}}{\sqrt{\Delta}-\sqrt{\UDelmOne}},
\end{equation}
where $\alpha_{\delta}=\big|\fdel^{-1}(\frac{1}{n})\big|$. The consequence of Assumption \ref{Assump1} in \eqref{eq:fdelLim} shows that for any $\epsilon>0$ and $0<x<1$, one can set $\delta$ sufficiently small such that $\big|\fdel^{-1}(x)\big|<\epsilon$. Therefore,
\begin{align*}
\lim_{\delta\rightarrow 0} \alpha_{\delta} = \lim_{\delta\rightarrow 0} \left|\fdel^{-1}\left(\frac{1}{n}\right)\right| = 0.
\end{align*}
This yields
\begin{align*}
\lim_{\delta\rightarrow 0} \|\Xb_0 - \Xb_{\delta}\|_F = 0.
\end{align*}
\end{proof}
\end{Thm4}

\section{Numerical Simulations} \label{sec:NumAna}
In this section, the performance of the SRF algorithm is evaluated empirically through simulations and is compared to {\color{\ROne_Col}some} other algorithms. In the first part of numerical experiments, effects of the algorithm parameters ($L, c,\text{and } \epsilon$) in reconstruction accuracy are studied. Next, in the second part, the so called \emph{phase transition} curve~\cite{RechFP10} between perfect recovery and failure is experimentally obtained for the SRF algorithm and is compared to that of the nuclear norm minimization. In the third part of simulations, accuracy and computational load of the SRF algorithm in solving the matrix completion problem are compared to {\color{\ROne_Col}five} well known matrix completion algorithms.

To generate a testing random matrix $\Xb \in \Rbb^{n_1 \times n_2}$ of rank $r$, the following procedure is used. We generate two random matrices $\Xb_L \in \Rbb^{n_1 \times r}$ and $\Xb_R \in \Rbb^{r \times n_2}$ whose entries are independent{\color{\FinEdt}ly} and identically drawn from a Gaussian distribution with zero mean and unit variance. Then $\Xb$ is constructed as the product of $\Xb_L$ and $\Xb_R$, i.e., $\Xb=\Xb_L\Xb_R$. {\color{\ROne_Col}Let $\Ab \in \Rbb^{m \times n_1 n_2}$ denote the matrix representation of $\ACal$ introduced in \eqref{MatRep}.} In the affine rank minimization problems, all entries of $\Ab$ are drawn independently and identically from a zero-mean, unit-variance Gaussian distribution. Moreover, in the matrix completion simulations, the index set $\Omega$ of revealed entries is selected uniformly at random. We denote the result of the SRF algorithm by $\Xbh$ and measure its accuracy by $\RSNR=20\log_{10}(\|\Xb\|_F / \|\Xb-\Xbh\|_F)$ in dB, which is referred to as the \emph{reconstruction SNR}. In addition, by term \emph{easy problems}, we mean problems in which the ratio $m/d_r$ is greater than 3, where $d_r=r(n_1+n_2-r)$ denotes the number of degrees of freedom in a real-valued rank--$r$ matrix~\cite{CandR09}. When this ratio is lower than or equal to 3, it is called a \emph{hard} problem.

In all experiments, the parameter $\mu$ is fixed at $1$, and we use a decreasing sequence of $\delta$'s according to $\delta_j=c\delta_{j-1},j\geq2$, where $0<c<1$ denotes the rate of decay. The value of $\delta_1$ is set twice as large as the largest singular value of the initial estimate. For the sake of simplicity, square matrices are tested, so $n_1=n_2=n$.

Our simulations are performed in MATLAB 8 environment using an Intel Core i7, 2.6 GHz processor with 8 GB of RAM, under Microsoft Windows 7 operating system.

\subsection{Parameters Effects}
\emph{Experiment 1.} As already discussed in Section~\ref{sec:mainidea}, it is not necessary to wait for complete convergence of the internal optimization loop. Instead, a few iterations suffice to only move toward the global maximizer for the current value of $\delta$. Thus, we suggested to do the internal loop for fixed $L$ times. However, the optimal choice of $L$ depends on the aspects of the problem at hand. As a rule of thumb, when the problem becomes harder, i.e., the number of measurements decreases toward the degrees of freedom, larger values of $L$ should be used. Likewise, for easier problems, smaller values of $L$ decrease the computational load of the algorithm, while the accuracy will not degrade very much.

\begin{figure}[tb]
\centering
\includegraphics[width=0.49\textwidth]{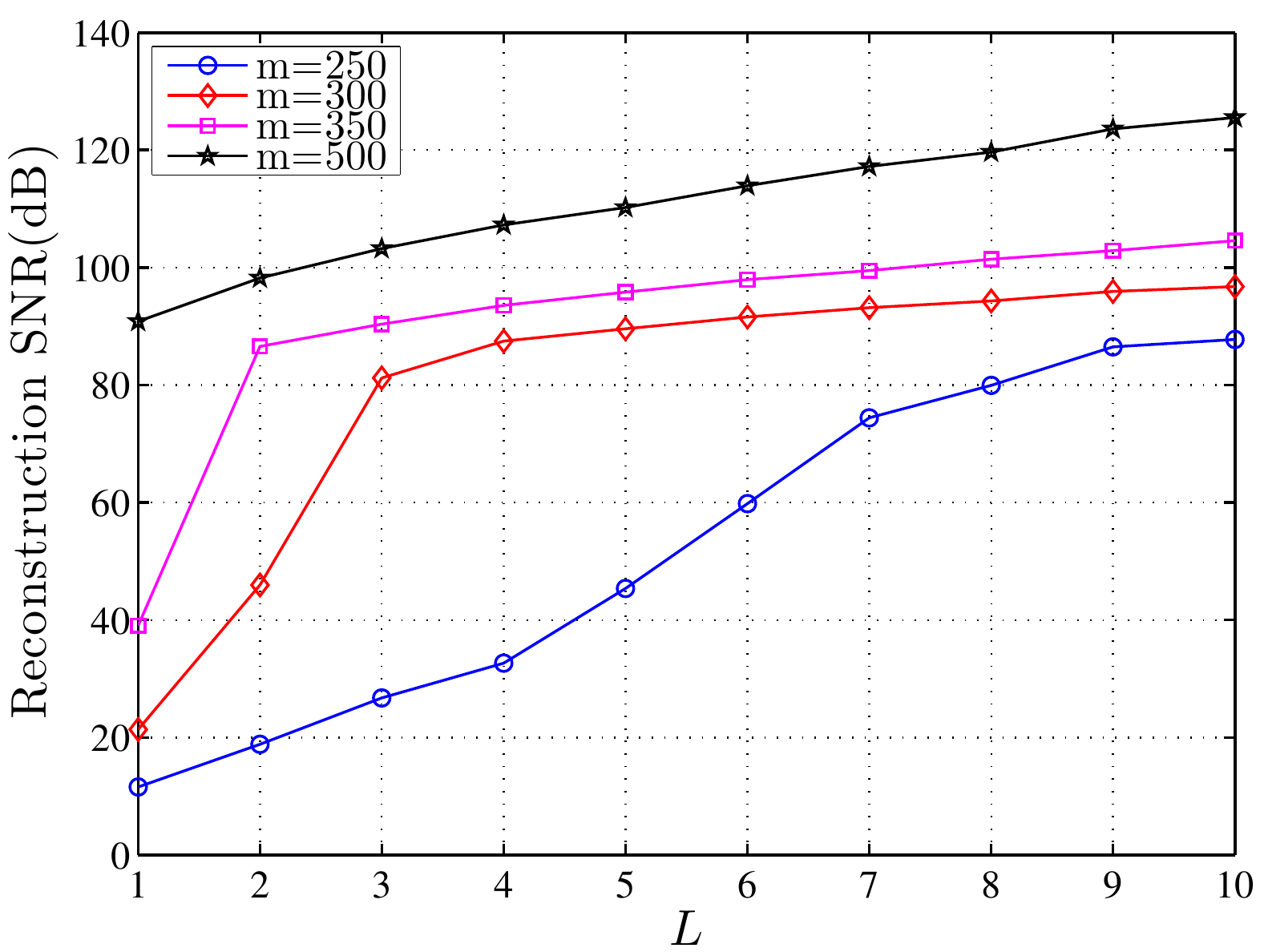}
\vspace{-0.6cm}
\caption{Averaged $\RSNR$ of the SRF algorithm in solving the ARM problem versus $L$. Matrix dimensions are fixed to $30 \times 30$, and $r$ is set to $3$. The parameter $c$ and $\epsilon$ are set to $0.9$ and $10^{-5}$, respectively, to have small effect on this analysis. SNR's are averaged over 100 runs.} \label{fig:LEff}
\end{figure}

To see the above rule, the affine rank minimization problem defined in~\eqref{RM} is solved using the SRF algorithm, while changing the parameter $L$. We put $n=30, r=3, \epsilon=10^{-5}$, and $c=0.9$. The number of measurements changes from 250 to 500 to cover both easy and hard problems. To obtain accurate $\RSNR$ estimates, the trials are repeated 100 times. Fig.~\ref{fig:LEff} shows the effects of changing $L$ from 1 to 10. It can be concluded from Fig.~\ref{fig:LEff} that for easy and hard problems, there is a threshold value for $L$, which choosing $L$ beyond it can only slightly improves reconstruction SNR. However, in {\color{\MinEdt}our} simulations, we found that increasing the $L$ boosts the computation time almost linearly. For instance, when $m=500$ and $L=1$, the average computation time is about 0.5 sec, while this time increases to about 1.2 sec for $L=5$ and to about 2.2 sec for $L=10$.

\begin{figure}[tb]
\centering
\includegraphics[width=0.49\textwidth]{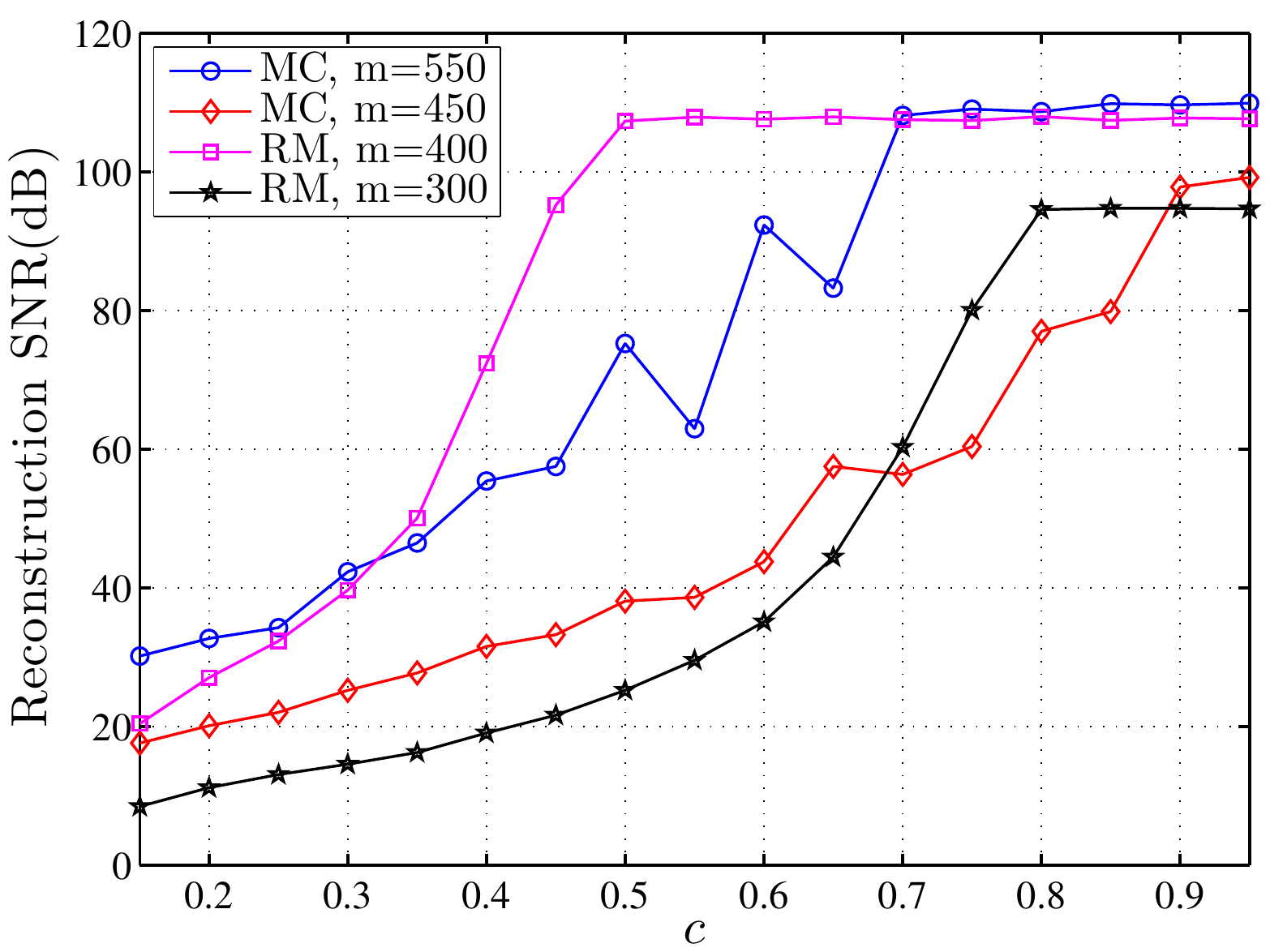}
\vspace{-0.6cm}
\caption{Averaged $\RSNR$ of the SRF algorithm as a function of $c$. Matrix dimensions are fixed to $30 \times 30$ and $r$ is set to $3$. The parameter $L$ and $\epsilon$ are set to $8$ and $10^{-5}$, respectively, to have small effect on this analysis. SNR's are averaged over 100 runs. `MC' and `RM' denote the matrix completion and affine rank minimization problems, respectively. For two MC problems, $m$ is set to $450$ and $550$, and for two RM problems, $m$ is set to $300$ and $400$.} \label{fig:cEff}
\end{figure}

\emph{Experiment 2.} The next experiment is devoted to the dependence of the accuracy of the SRF algorithm on the parameter $c$. In this experiment, the dimensions of the matrix are the same as the in previous experiment, and $L$ and $\epsilon$ are fixed to $8$ and $10^{-5}$, respectively. Affine rank minimization and matrix completion problems are solved with two different number of measurements to show the effect on different conditions. $c$ is changed from 0.15 to 0.95, $\RSNR$'s are averaged on 100 runs. Fig.~\ref{fig:cEff} depicts the reconstruction SNR versus the parameter $c$ for different problems. It is obvious that SNR increases as $c$ approaches 1. However, when $c$ exceeds a critical value, SNR remains almost constant.

Generally, the optimal choice of $c$ depends on the criterion which aimed to be optimized. When accuracy is the key criterion, $c$ should be chosen close to 1, which results in slow decay in the sequence of $\delta$ and a higher computational time.

\begin{figure}[tb]
\centering
\includegraphics[width=0.49\textwidth]{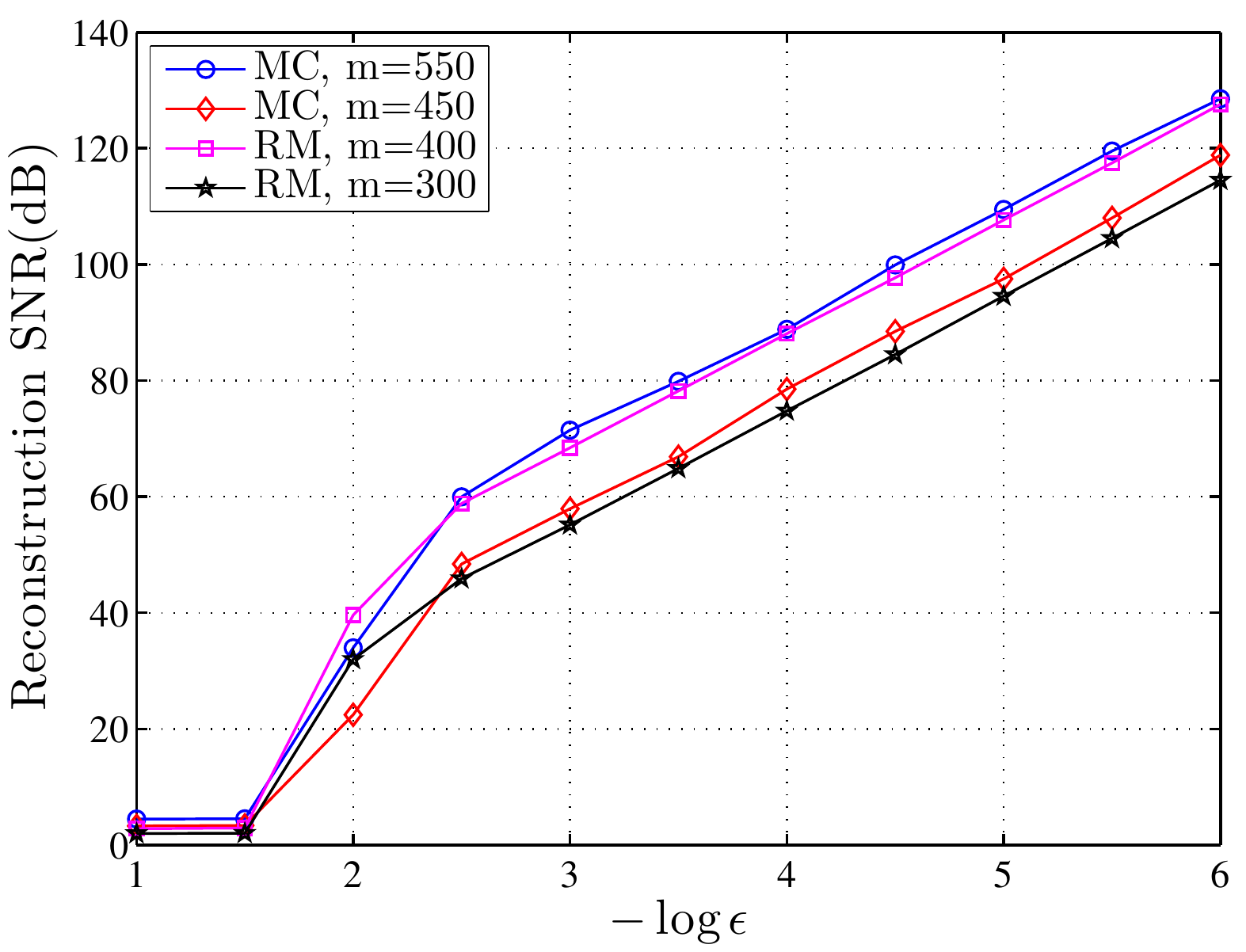}
\vspace{-0.6cm}
\caption{Averaged $\RSNR$ of the SRF algorithm as a function of $\epsilon$. Matrix dimensions are fixed to $30 \times 30$ and $r$ is set to $3$. The parameter $L$ and $c$ are set to 8 and 0.9, respectively, to have small effect on this analysis. SNR's are averaged over 100 runs. $\epsilon$ is changed from $10^{-1}$ to $10^{-6}$. `MC' and `RM' denote the matrix completion and affine rank minimization problems, respectively. For two MC problems, $m$ is set to $450$ and $550$, and for two RM problems, $m$ is set to $300$ and $400$.} \label{fig:EpEff}
\end{figure}

\emph{Experiment 3.} In this experiment, the effect of $\epsilon$ on the accuracy of the algorithm is analyzed. All dimensions and parameters are the same as in the experiment 2 except $c$ and $\epsilon$. $c$ is fixed to $0.9$, and $\epsilon$ is changed from $10^{-1}$ to $10^{-6}$. The result of this experiment is shown in Fig.~\ref{fig:EpEff}. It is seen that after passing a critical value, logarithmic reconstruction SNR increases almost linearly as $\epsilon$ decreases linearly in logarithmic scale. Hence, it can be concluded that $\epsilon$ controls the closeness of the final solution to the minimum rank solution.

\subsection{Phase Transition Curve}
\emph{Experiment 4.} To the best of our knowledge, the tightest available bound on the number of required samples for the NNM to find the minimum rank solution is two times greater than that of the rank minimization problem~\cite{MohaFH11}. More precisely, for the given linear operator which has a null space with the $\Delta$-spherical section property,~\eqref{RM} has a unique solution if $\rank(\Xb_0) < \Delta/2$, while \eqref{NM} and~\eqref{RM} share a common solution if $\rank(\Xb_0) < \Delta/4$. Our main goal, in this experiment, is to show that the SRF algorithm can recover the solution in situations where nuclear norm minimization fails. In other words, this algorithm can get closer to the intrinsic bound in recovery of low-rank matrices. The computational cost of the SRF algorithm will be compared to efficient implementations of the nuclear norm minimization in the next experiment.

\begin{figure}[tb]
\centering
\includegraphics[width=0.49\textwidth]{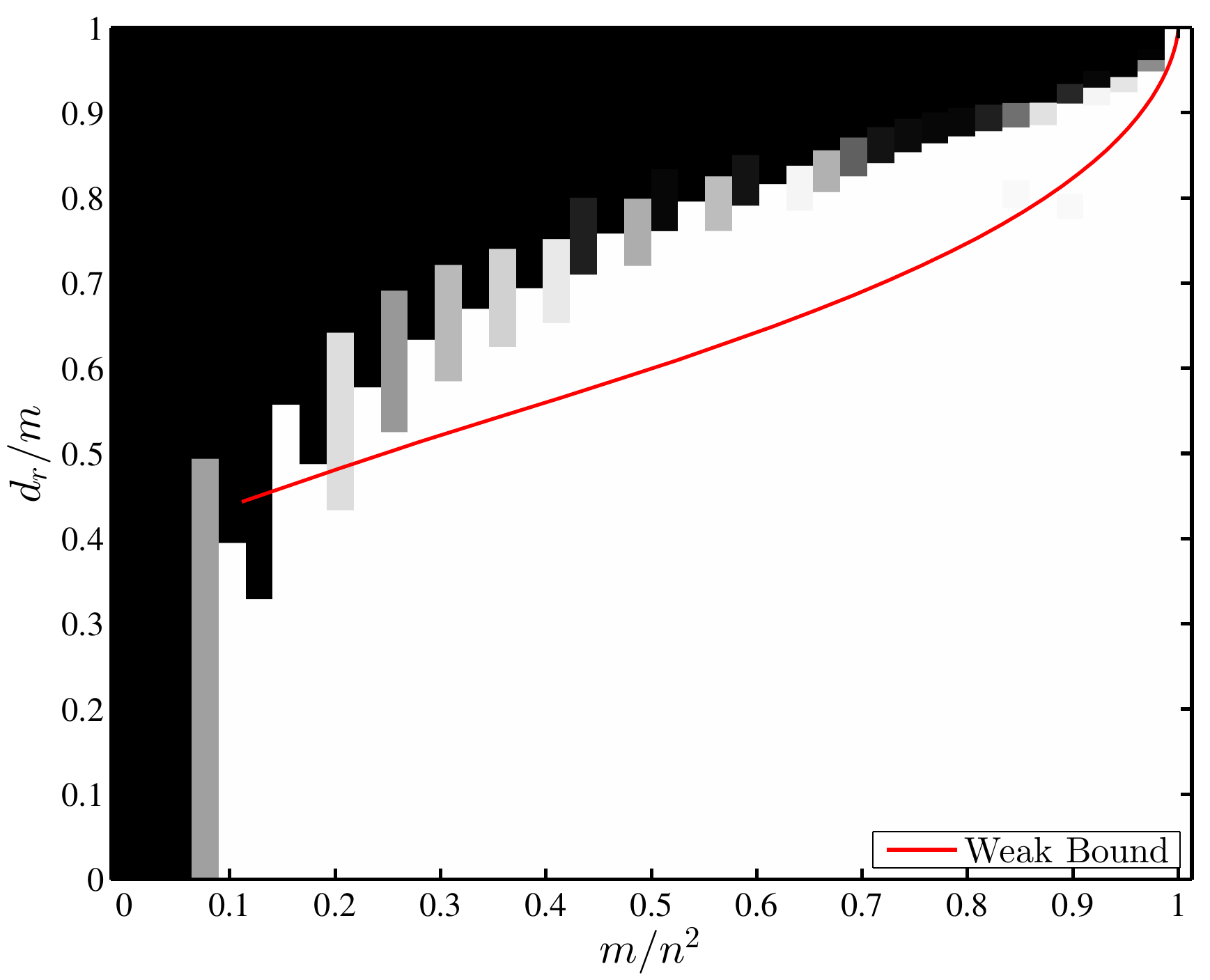}
\vspace{-0.6cm}
\caption{Phase transition of the SRF algorithm in solving the ARM problem. $n=40, \epsilon=10^{-5}, L=6, c=0.9$, and simulations are performed 50 times. Gray-scale  color of each cell indicates the rate of perfect recovery. White denotes 100\% recovery rate, and black denotes 0\% recovery rate. A recovery is perfect if the $\RSNR$ is greater than 60 dB. The red trace shows the so called weak bound derived in~\cite{OymaH10} for the number of required measurements for perfect recovery of low-rank matrix using the nuclear norm heuristics.} \label{fig:FigPhCurve}
\end{figure}

Like compressive sensing literature, the phase transition can be used to indicate the region of perfect recovery and failure~\cite{RechFP10}. Fig.~\ref{fig:FigPhCurve} shows the results of applying the proposed algorithm on the affine rank minimization. A solution is declared to be recovered if reconstruction SNR is greater than 60 dB. The matrix dimension is $40 \times 40, \epsilon=10^{-5}, L=6, \text{ and } c=0.9$. Simulations are repeated 50 times. The gray color of cells indicates the empirical recovery rate. White denotes perfect recovery in all trials, and black shows unsuccessful recovery for all experiments. Furthermore, the thin trace on the figure shows a theoretical bound in recovery of low-rank solutions via the nuclear norm minimization found in~\cite{OymaH10}. In~\cite{OymaH10}, it is shown that this bound is very consistent to the numerical simulations; thus, we use it for the sake of comparison. One can see in Fig.~\ref{fig:FigPhCurve} that there is a very clear gap between this bound and phase transition of the SRF algorithm.

\subsection{Matrix Completion}
\emph{Experiment 5.} The accuracy and computational costs of the proposed algorithm in solving the matrix completion problem are analyzed and compared to {\color{\ROne_Col}five} other methods. Among many available approaches, {\color{\ROne_Col}IALM \cite{LinChM09}, APG \cite{TohY10}, LMaFit \cite{WenYZ12}, BiG-AMP \cite{ParkSC13},} and OptSpace~\cite{KeshMO10} are selected as competitors. {\color{\ROne_Col}IALM and APG are} efficient implementations of the NNM and can obtain very accurate results with low complexity~\cite{LinChM09,TohY10}, while {\color{\ROne_Col}other selected methods are only applicable to the MC setting and exploit other heuristics rather than the nuclear norm to find a low-rank solution. LMaFit, which is known to be very fast in completing partially observed matrices, uses a nonlinear successive over-relaxation algorithm \cite{WenYZ12}. BiG-AMP extends the generalized approximate message passing algorithm in the compressive sensing to the matrix completion and outperforms many state-of-the-art algorithms \cite{ParkSC13}.} OptSpace is based on trimming rows and columns of the incomplete matrix followed by truncation of some singular values of the trimmed matrix~\cite{KeshMO10}.

{\color{\ROne_Col} LMaFit, BiG-AMP, and OptSpace require an accurate estimate of the rank of the solution. MATLAB implementation of OptSpace\footnote{{\color{\ROne_Col}MATLAB code: web.engr.illinois.edu/$_{\widetilde{~}}$swoh/software/optspace/code.html}} is provided with a function for estimating the rank of the solution, and we use it in running OptSpace. Moreover, LMaFit\footnote{{\color{\ROne_Col}MATLAB code: lmafit.blogs.rice.edu/}} should be initialized with an upperbound on the rank of the solution which, in our numerical experiments, this upperbound is set to $\frac{1}{2} n$. Also, BiG-AMP\footnote{{\color{\ROne_Col}MATLAB code: sourceforge.net/projects/gampmatlab/}} needs a similar upperbound to learn the underlying rank, and we pass $\frac{1}{2} n$ as the upperbound to the Big-AMP algorithm too.}

\begin{figure*}[ht!]
        \centering
        \subfigure{%
                \label{fig:MC_SNR1}
                \includegraphics[width=0.32\textwidth]{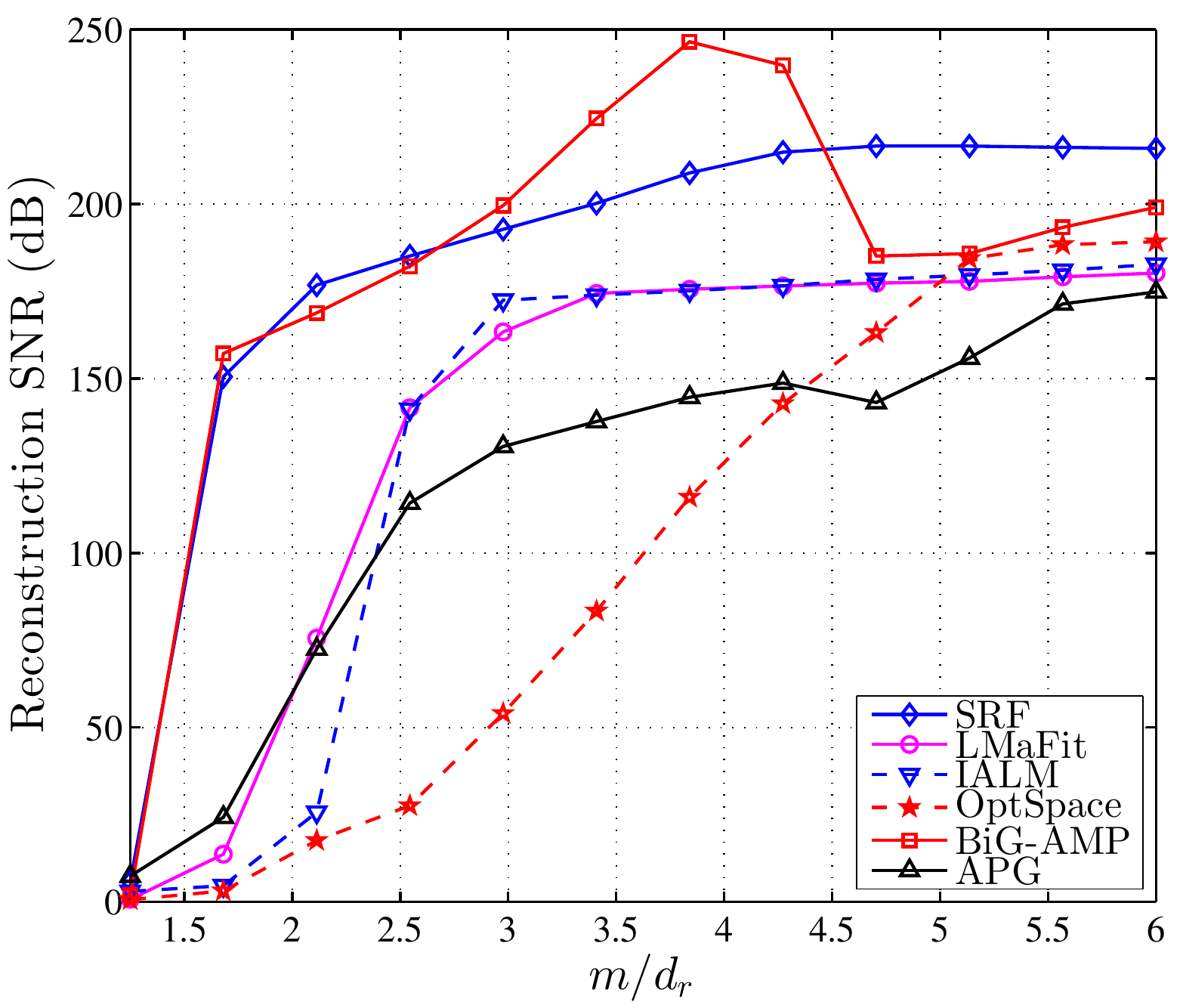}
        }%
        \subfigure{%
                \label{fig:MC_SNR2}
                \includegraphics[width=0.32\textwidth]{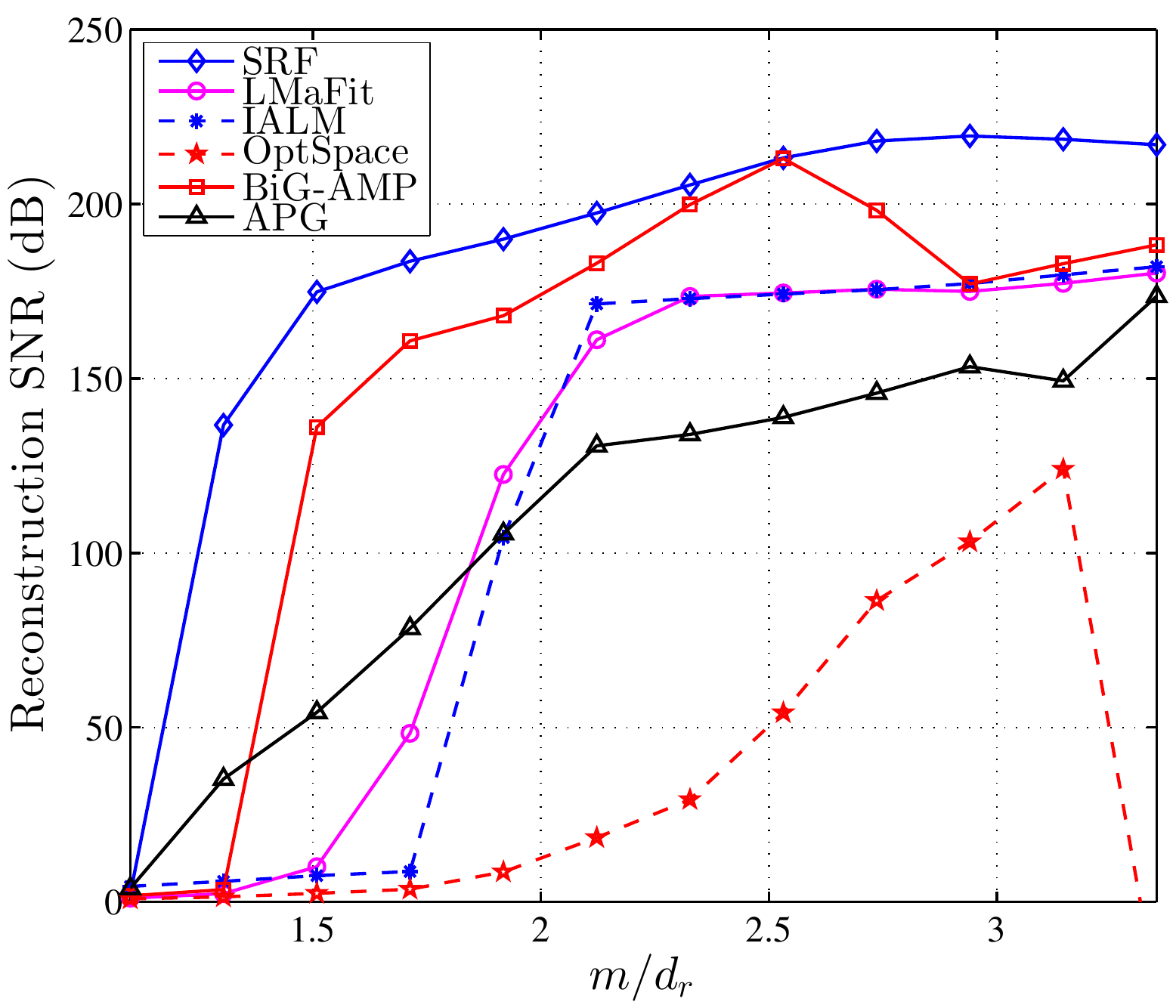}
        }%
        \subfigure{%
                \label{fig:MC_SNR3}
                \includegraphics[width=0.32\textwidth]{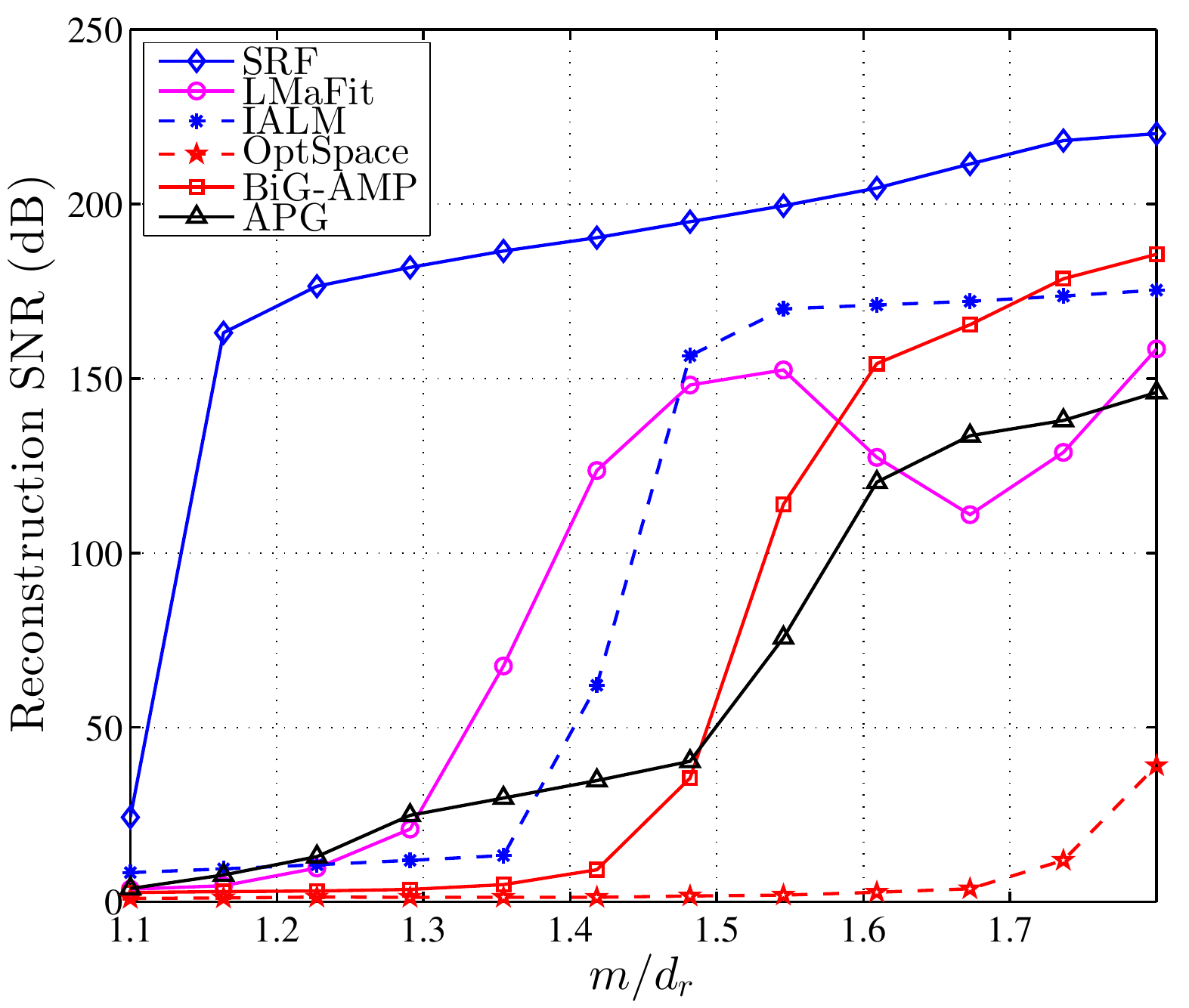}
        }\addtocounter{subfigure}{-3}%
        \vspace{-0.4cm}
        \subfigure[$r = 8$.]{%
                \label{fig:MC_Time1}
                \includegraphics[width=0.32\textwidth]{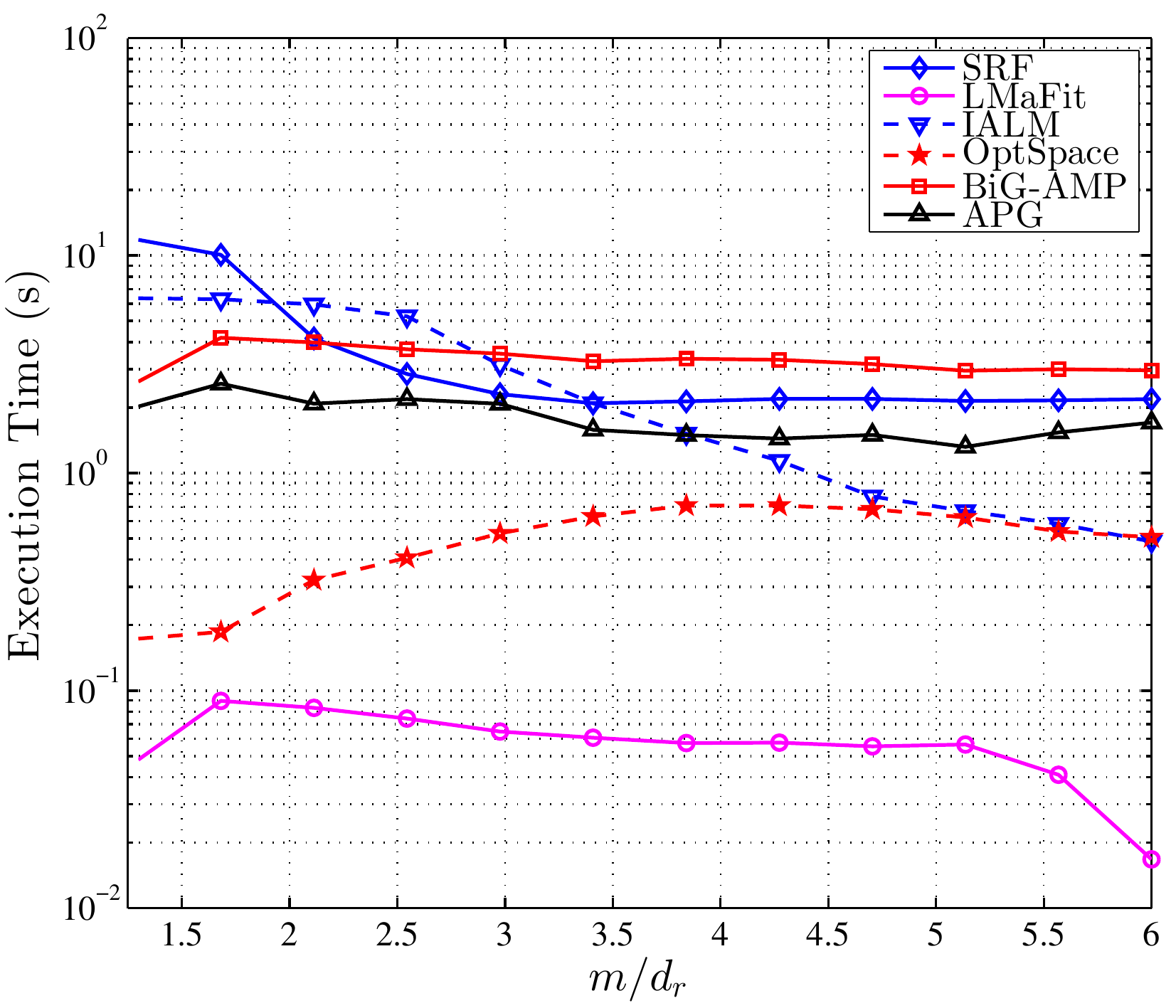}
        }%
        \subfigure[$r = 16$.]{%
                \label{fig:MC_Time2}
                \includegraphics[width=0.32\textwidth]{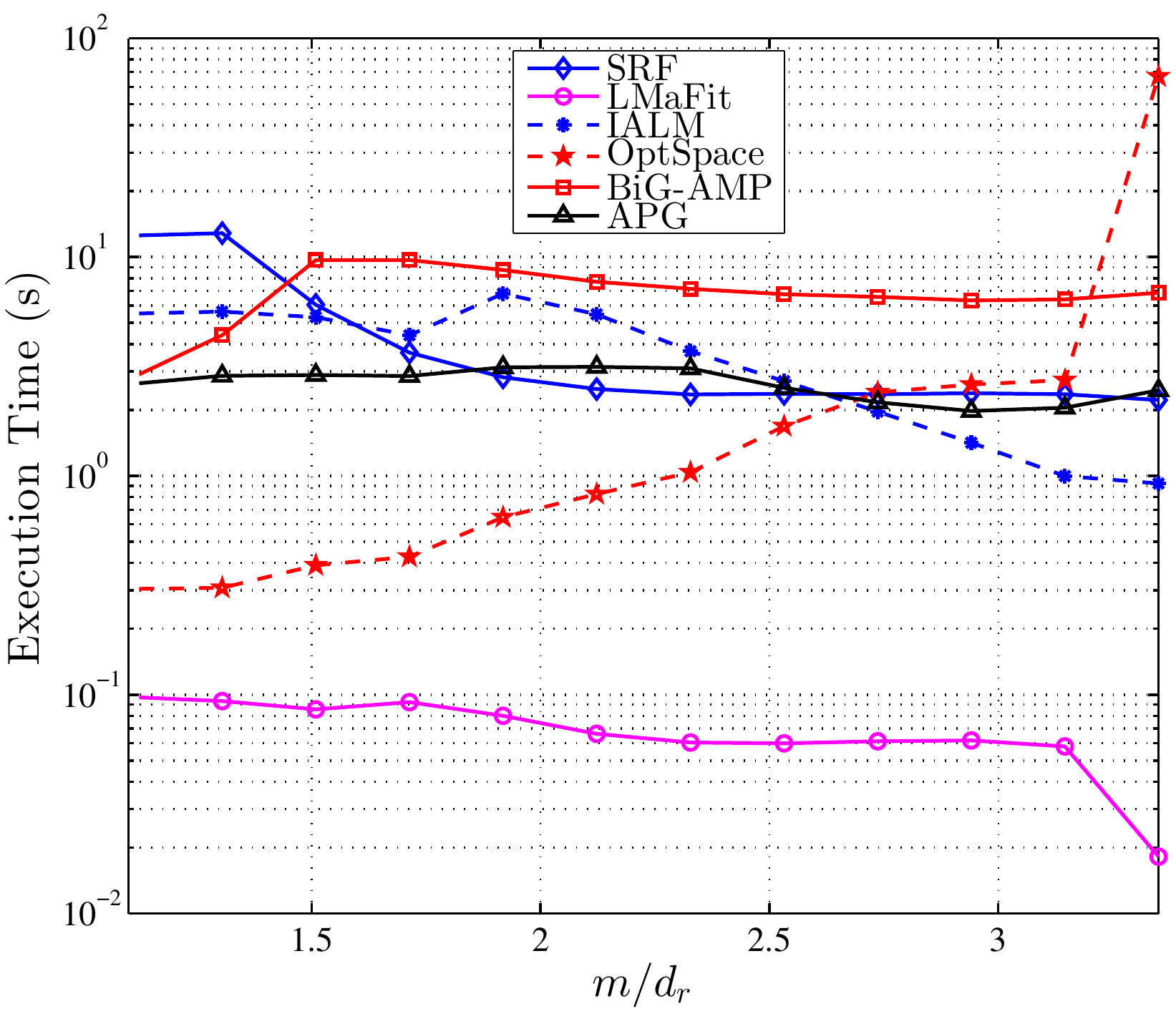}
        }%
        \subfigure[$r = 32$.]{%
                \label{fig:MCCompFig3}
                \includegraphics[width=0.32\textwidth]{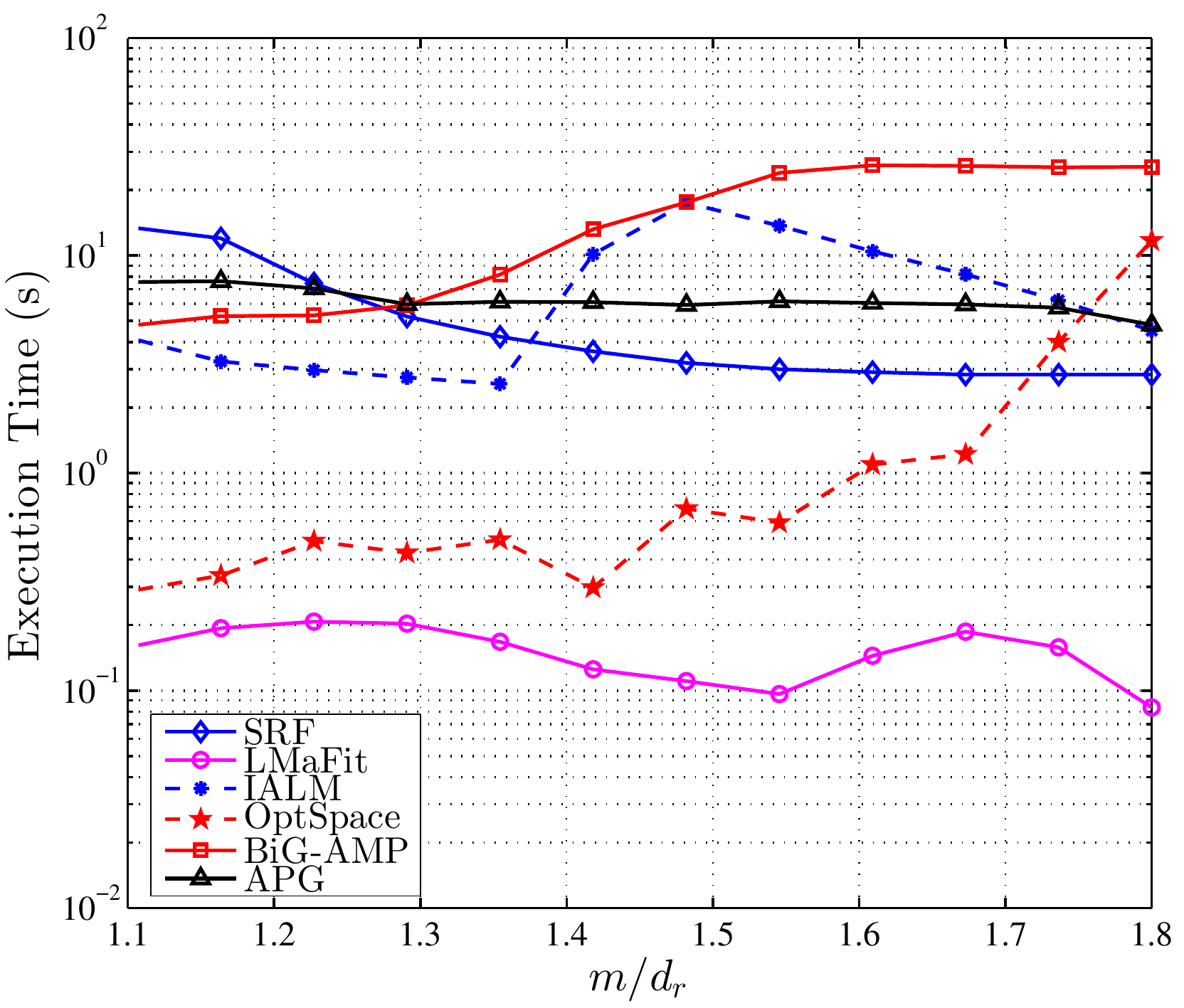}
        }%
        \vspace{-0.2cm}
        \caption{{\color{\ROne_Col}Comparison of the SRF algorithm with the IALM \cite{LinChM09}, APG \cite{TohY10}, LMaFit \cite{WenYZ12}, BiG-AMP \cite{ParkSC13}, and OptSpace~\cite{KeshMO10} algorithms in terms of accuracy and execution time in completing low-rank matrices. Averaged $\RSNR$ and execution time of all algorithm are plotted as a function of $m/d_r$. Matrix dimensions are fixed to $100 \times 100$, and $r$ is set to 8, 16, and 32. Trials are repeated 100 times, and results are averaged.}}\label{fig:MCCompFig}
\end{figure*}

{\color{\ROne_Col}IALM\footnote{{\color{\ROne_Col}MATLAB code: perception.csl.illinois.edu/matrix-rank/sample\_code.html}}, LMaFit, and OptSpace are run by their default parameters except for $\texttt{tol}=10^{-9}$. For APG\footnote{{\color{\ROne_Col}MATLAB code: math.nus.edu.sg/$_{\widetilde{~}}$mattohkc/NNLS.html}}, we use default parameters and set \texttt{tol} and \texttt{mu\_scaling} to $10^{-9}$ to have the best achieved $\RSNR$ on the same order of other methods. SRF is run with $\epsilon=10^{-9},L=8,\text{ and } c=0.95.$}

{\color{\ROne_Col}Matrix dimensions are fixed to $100 \times 100$, and $r$ is set to 8, 16, and 32. To see the performance of the aforementioned algorithms, $\RSNR$ and execution time are reported as a function of $m/d_r$ for the three values of the rank.} Although CPU time is not an accurate measure of the computational costs, we use it as a rough estimate to compare algorithm complexities. Every simulation is run $100$ times, and the results are averaged.

{\color{\ROne_Col}Fig. \ref{fig:MCCompFig} demonstrates the results of these comparisons for the three matrix ranks as a function of number of measurements. In comparison to BiG-AMP, while SRF starts completing low-rank matrices with a good accuracy approximately with the same number of measurements when the rank equals to 8, once $r$ increases to 16, it needs smaller number of measurements to successfully recover the solutions. This gap is widen when $r=32$. Furthermore, in all simulated cases, SRF has lower running time when compared to BiG-AMP except for starting values of $m/d_r$. SRF also outperforms IALM and APG, which implement NNM, in terms of accuracy, whereas its computational complexity is very close to that of APG. Finally, although the execution time of LMaFit is considerably lower than that of SRF, it needs much larger number of measurements to start recovering low-rank solutions.} Note that, here, $c$ is set to 0.95 to accommodate the worst case scenario of hard problems. However, it can be tuned to speed up the SRF method, if the working regime is \textit{a priori} known.

{\color{\ROne_Col}In summary, the significant advantage of SRF is in solving hard problems where the number of measurements is approaching to $d_r$. Especially, when the matrix rank increases (see Fig. \ref{fig:MC_Time2} and \ref{fig:MCCompFig3}), SRF can recover the low-rank solution with at least 20\% less number of measurements than other competitors.}

\section{Conclusion} \label{sec:Con}
In this work, a rank minimization technique based on approximating the rank function and successively improving the quality of the  approximation was proposed. We theoretically showed that the proposed iterative method asymptotically achieves the solution to the rank minimization problem, provided that the middle-stage minimizations are exact. We further examined the performance of this method using numerical simulations. The comparisons against {\color{\MinEdt}five} common methods reveal superiority of the proposed technique in terms of {\color{\MinEdt}accuracy}, especially when the number of affine measurements decreases toward the unique representation lower-bound. By providing examples, we even demonstrate the existence of scenarios in which the conventional nuclear norm minimization fails to recover the unique low-rank matrix associated with the linear constraints, while the proposed method succeeds.

\appendices
\section{} \label{appA}
In this appendix, the closed form least-squares solution of the orthogonal back projection onto the feasible set is derived. Let us cast the affine constraints $\ACal(\Xb)=\bb$ as $\mathbf{A}\vect(\Xb)=\bb$. The goal is to find the nearest point in the affine set to the result of the $j$-th iteration, $\Xb_j$. Mathematically,
\begin{equation} \label{LSMat}
\min_{\Xb} \|\Xb - \Xb_j\|_F^2 \text{ subject to } \ACal(\Xb)=\bb,
\end{equation}
or equivalently,
\begin{equation} \label{LSVect}
\min_{\Xb} \|\vect(\Xb) - \vect(\Xb_j)\|^2 \text{ subject to } \mathbf{A}\vect(\Xb)=\bb,
\end{equation}
where $\|\cdot\|$ denotes vector $\ell_2$-norm.
By putting $\mathbf{y} = \vect(\Xb) - \vect(\Xb_j)$, the problem~\eqref{LSVect} can be easily cast as the following least-squares problem
\begin{equation*}
\min_{\mathbf{y}} \|\mathbf{y}\|_2^2 \text{ subject to } \mathbf{Ay} = \bb - \mathbf{A}\vect(\Xb_j).
\end{equation*}
Let $\mathbf{A}^\dagger = \mathbf{A}^T(\mathbf{A}\mathbf{A}^T)^{-1}$ be the Moore-–Penrose pseudoinverse of $\mathbf{A}$. Then the least-squares solution of~\eqref{LSMat} will be $\Xb = \mat_{n_1,n_2}\big(\mathbf{A}^\dagger \bb + [\mathbf{I} - \mathbf{A}^\dagger \mathbf{A} ] \vect({\Xb_j})\big)$, where $\mathbf{I}$ denotes the identity matrix, and $\mat_{n_1,n_2}(\cdot)$ reverses the operation of vectorization, i.e., $\mat_{n_1,n_2}\big(\vect(\Xb)\big)=\Xb$.

\section{Proof of Theorem~\ref{InitThm}} \label{appB}
\begin{proof}
Let $\Xb_{\delta}=\argmax \{\Fdel(\Xb) \;|\; \ACal(\Xb)=\bb\}$. To prove $\lim_{\delta \to \infty} \Xb_{\delta}=\XbF$, we first focus on singular values $\sigma_i(\Xbd)$. Due to Assumption \ref{Assump1}, it is known that $\lim_{\delta\rightarrow \infty} \Fdel(\XbF) = n$. Thus, for any $\epsilon\geq 0$, one can set $\delta$ large enough such that $\Fdel(\XbF)\geq n-\epsilon$. Note that for any $1 \leq i \leq n$, we have that
\begin{align*}
n-1+\fdel\big(\sigma_i(\Xbd)\big) \geq \Fdel(\Xb_{\delta}) \geq \Fdel(\XbF)\geq n-\epsilon,
\end{align*}
or
\begin{align*}
\fdel\big(\sigma_i(\Xbd)\big) \geq 1-\epsilon.
\end{align*}
This implies that $\sigma_i(\Xbd) \leq |\fdel^{-1}(1-\epsilon)| = \delta |f^{-1}(1-\epsilon)|$. Hence,
\begin{align*}
0\leq \lim_{\delta\rightarrow\infty} \frac{\sigma_i(\Xbd)}{\delta} \leq \big|f^{-1}(1-\epsilon)\big| , ~~~ \forall~0<\epsilon<1.
\end{align*}
By considering the above inequality for $\epsilon\rightarrow 0$, we conclude that
\begin{align*}
\lim_{\delta\rightarrow\infty} \frac{\sigma_i(\Xbd)}{\delta} = 0,~~~ 1 \leq i \leq n.
\end{align*}

Using the Taylor expansion, we can rewrite $f(\cdot)$ as
\begin{equation*}
f(s)=1-\gamma s^2+g(s),
\end{equation*}
where $\gamma=-\frac{1}{2}f''(0)$ and
\begin{equation} \label{glim}
\lim_{s \to 0} \frac{g(s)}{s^2}=0.
\end{equation}
In turn, $\Fdel(\cdot)$ can be rewritten as
\begin{IEEEeqnarray}{rCl} \label{FTaylor}
\Fdel(\Xb) & = & \sumn \fdel\big(\sigma_i(\Xb)\big) \nonumber \\
& = &n-\frac{\gamma}{\delta^2}\sumn \sigma_i^2(\Xb) + \sumn g(\sigma_i(\Xb)/\delta).
\end{IEEEeqnarray}
This helps us rewrite $\Fdel(\Xbd) \geq \Fdel(\XbF)$ in the form
\begin{multline*}
\frac{\gamma}{\delta^2}\sumn \sigma_i^2(\Xbd) - \sumn g(\sigma_i(\Xbd)/\delta) \leq \\
\frac{\gamma}{\delta^2}\sumn \sigma_i^2(\XbF) - \sumn g(\sigma_i(\XbF)/\delta),
\end{multline*}
or similarly,
\begin{align*}
\|\SigbXbd\|^2 - \|\SigbXbF\|^2 \leq & \frac{\sumn g\big(\sigma_i(\Xbd)/\delta\big) - g\big(\sigma_i(\XbF)/\delta\big)}{\gamma \, \delta^{-2}}  \\
\leq & \phantom{+}\frac{\|\SigbXbd\|^2}{\gamma} \sumn \frac{\big|g\big(\sigma_i(\Xbd)/\delta\big)\big|}{\big(\sigma_i(\Xbd)/\delta\big)^2} \\
& + \frac{\|\SigbXbF\|^2}{\gamma} \sumn \frac{\big|g\big(\sigma_i(\XbF)/\delta\big)\big|}{\big(\sigma_i(\XbF)/\delta\big)^2}.
\end{align*}
Recalling $\|\SigbXb\|^2=\|\Xb\|_F^2$, we can write that
\begin{equation} \label{lim1}
\|\Xbd\|_F^2 \leq \|\XbF\|_F^2 \frac{1 + \frac{1}{\gamma} \bigg(\sumn \Big| \frac{g\big(\sigma_i(\XbF)/\delta\big)}{\big(\sigma_i(\XbF)/\delta\big)^2} \Big|\bigg)}{\bigg|1 - \frac{1}{\gamma} \bigg(\sumn \Big| \frac{g\big(\sigma_i(\Xbd)/\delta\big)}{\big(\sigma_i(\Xbd)/\delta\big)^2} \Big|\bigg)\bigg|}.
\end{equation}
We also have
\begin{eqnarray} \label{lim2}
\lim_{\delta \to \infty} \sigma_i(\XbF)/\delta = 0 & \xRightarrow{~\eqref{glim}}   \lim_{\delta \to \infty} \frac{g\big(\sigma_i(\XbF)/\delta\big)}{\big(\sigma_i(\XbF)/\delta\big)^2} = 0, \\
\lim_{\delta \to \infty} \sigma_i(\Xbd)/\delta = 0 & \xRightarrow{~\eqref{glim}}   \lim_{\delta \to \infty} \frac{g\big(\sigma_i(\Xbd)/\delta\big)}{\big(\sigma_i(\Xbd)/\delta\big)^2} = 0. \label{lim3}
\end{eqnarray}
Application of \eqref{lim2} and \eqref{lim3} in \eqref{lim1} results in
\begin{equation} \label{lim4}
\lim_{\delta \to \infty} \|\Xbd\|_F^2 \leq \|\XbF\|_F^2.
\end{equation}

According to the definition of $\XbF$, we have $\|\Xbd\|_F^2 \geq \|\XbF\|_F^2$ and $\lim_{\delta \to \infty} \|\Xbd\|_F^2 \geq \|\XbF\|_F^2$. Combining this result with~\eqref{lim4}, we obtain
\begin{equation*}
\lim_{\delta \to \infty} \|\Xbd\|_F^2 = \|\XbF\|_F^2.
\end{equation*}

Also, any matrix in $\nullS(\ACal)$ is perpendicular to $\XbF$ since it is the minimum Frobenius norm solution of the $\ACal(\Xb)=\bb$. {\color{\RTre_Col}To see this, let $\ACal^*:\Rbb^{m} \to \Rbb^{n_1 \times n_2}$ denote the adjoint operator of $\ACal$ and let $\mathcal{B}:\Rbb^{m} \to \Rbb^{m}$ denote the inverse of the operator $\ACal(\ACal^*(\cdot))$. Then, similar to the vector case, one can show that $\XbF = \ACal^*(\mathcal{B}(\bb))$ and $\forall \Zb \in \nullS(\ACal), \langle \XbF, \Zb \rangle = \trace(\XbF^T \Zb) = 0$}. Thus,
\begin{equation*}
\|\Xbd\|_F^2 = \|\XbF\|_F^2 + \|\Xbd-\XbF\|_F^2.
\end{equation*}

In summary, we conclude that $\lim_{\delta \to \infty} \|\Xbd-\XbF\|_F^2 = 0$ which establishes $\lim_{\delta \to \infty} \Xbd=\XbF$.
\end{proof}

\section*{Acknowledgment}
\addcontentsline{toc}{section}{Acknowledgment}
The authors would like to thank Hooshang Ghasemi for his help in obtaining preliminary results {\color{\MinEdt}and anonymous reviewers for their helpful comments.}

\bibliography{SepSrc}
\bibliographystyle{IEEEbib}

\end{document}